\documentclass[a4paper,fleqn]{cas-dc}

\usepackage[numbers]{natbib}
\usepackage{hyperref}               
\usepackage{cleveref}
\usepackage{subcaption}
\usepackage{algorithm}
\usepackage{algpseudocode}
\usepackage{amsthm}

\theoremstyle{definition}
\newtheorem{definition}{Definition}

\theoremstyle{remark}
\newtheorem*{remark}{Remark}

\def\tsc#1{\csdef{#1}{\textsc{\lowercase{#1}}\xspace}}
\tsc{WGM}
\tsc{QE}
\tsc{EP}
\tsc{PMS}
\tsc{BEC}
\tsc{DE}

\begin{document}
\let\WriteBookmarks\relax
\def\floatpagepagefraction{1}
\def\textpagefraction{.001}
\shorttitle{The Social Sphere Model}
\shortauthors{Lin, Schaposnik, Wu}

\title [mode = title]{The Social Sphere Model: Heuristic Influence Prediction in Evolving Networks}

\author[1]{Marina Lin}
\credit{Formal analysis, Investigation, Writing - Review \& Editing}

\author[2]{Laura P. Schaposnik}
\credit{Supervision, Conceptualization, Writing - Review \& Editing, Funding acquisition, Project administration}
\author[3]{Raina Wu}
\credit{Conceptualization, Methodology, Software, Formal analysis, Investigation, Writing - Original Draft, Visualization}

\affiliation[1]{organization={Harvard University}, 
                city={Cambridge}, 
                state={MA}, 
                country={USA}}

\affiliation[2]{organization={University of Illinois at Chicago}, 
                city={Chicago}, 
                state={IL}, 
                country={USA}}

\affiliation[3]{organization={MIT}, 
                city={Cambridge}, 
                state={MA}, 
                country={USA}}

\cortext[cor1]{Corresponding author. Email: \href{mailto:schapos@uic.edu}{schapos@uic.edu}}

\fntext[1]{Contributing authors: \href{mailto:marinalin@college.harvard.edu}{marinalin@college.harvard.edu}, \href{mailto:rwu2024@gmail.com}{rwu2024@gmail.com}}

\begin{abstract}
 How would admissions look like in a {\it university program for influencers}? In the realm of social network analysis, influence maximization and link prediction stand out as pivotal challenges. Influence maximization focuses on identifying a set of key nodes to maximize information dissemination, while link prediction aims to foresee potential connections within the network. These strategies, primarily deep learning link prediction methods and greedy algorithms, have been previously used in tandem to identify future influencers. However, given the complexity of these tasks, especially in large-scale networks, we propose an algorithm, {\bf The Social Sphere Model}, which uniquely utilizes expected value in its future graph prediction and combines specifically path-based link prediction metrics and heuristic influence maximization strategies to effectively identify future vital nodes in weighted networks. Our approach is tested on two distinct contagion models, offering a promising solution with lower computational demands. This advancement not only enhances our understanding of network dynamics but also opens new avenues for efficient network management and influence strategy development.
\end{abstract}

\maketitle

\section{Introduction}

 \label{sec:intro}
When a company is launching a new product and wants to use social media for marketing, understanding who to hire for such campaigns becomes of most importance to maximize profit. With the explosion of social media in recent years, a common strategy has been to target certain influential individuals for advertising. These individuals, or ``influencers," become the initial spreaders in different social settings, and through their connections and followers,  companies reach greater awareness (and sales) in the entire society. However, social groups are constantly evolving and thus influencers in certain social setting may no longer be the optimal selection in the few months or years until the product is released, which can lower efficiency. As a result, being able to predict future influencers is extremely important.

Within a mathematical setting, one can consider social networks as graphs, where the nodes represent members of the network, and their connections are represented by edges. The study of how nodes can influence a network has been done for many decades, and the reader can refer to recent surveys such as Lü et al.'s \cite{listofidentificationtypes} in 2016, Pei et al.'s \cite{survey2} in 2019, and AbdulAmeer et al.'s \cite{survey1} in 2022.

In particular,  Domingos and Richardson \cite{imp1} studied influence through a viral marketing perspective. In 2003, Kempe et al. \cite{imp2} formalized this as the influence maximization problem (IMP) of choosing $k$ nodes of greatest influence, proving that a complete solution was NP-hard. However, it is interesting to note that to the best of our knowledge, no mathematical definition of a social \textit{\textbf{Influencer}} as been proposed in the literature, and thus we shall introduce here what we think can be a useful way of thinking of such important individuals within our society (see Definition \ref{influencer}). 

In the present paper we will focus on heuristic measures, revolving around local centralities due to lower time complexity: while global metrics take into consideration more complete information about the graph, they are often computationally expensive. Furthermore, Hu et al. \cite{hu} claimed that global influence can be approximated by local influence from neighbors of order around three to four, which means local measures may be more cost-effective. As mentioned before, one should keep in mind that social networks are constantly changing. In 2004, Liben-Nowell and Kleinberg \cite{linkpredictioncomparison} formalized the link prediction problem for social networks as predicting future edges based on knowledge of the current graph. Hasan and Zaki \cite{linkpredictionsurvey} and Lü and Zhou \cite{linkpredictionsurvey2} provided surveys of link prediction results. In our work, we shall focus on local similarity metrics and their quasi-local extensions given by Aziz et al. \cite{globalextensions}, which are heuristic measures similar to centrality.

In 2019, Ghafouri and Khasteh \cite{lpim1} introduced the idea of using a graph modelling technique to engage in link prediction prior to use of greedy algorithms for influence maximization as a way to account from invisible edges between nodes (i.e. edges not observed when collecting data on the network). Singh and Kailasam \cite{lpim2} utilized a similar approach when predicting sets of influential nodes for the next iteration of the graph through link prediction using the a type of Restricted Boltzmann Machine (RBM). In 2023, Yanchenko et al. \cite{exante} discussed the use of deep learning link prediction to understand the outcomes of the future networks, then identify influential nodes through the greedy algorithm and dynamic degree discount. We take a similar approach, but focus on heuristic link prediction measures whereas they consider linear regression, deep learning, matrix factorization, and graph neural networks, and more. We also evaluate many other common centrality metrics and algorithms that, to the best of our knowledge, previous literature has not covered. In an effort to lower time complexity, our graph prediction also makes use of expected value to decrease computation cost of influence spread.

We seeks to fill a very important gap in the literature which would allow the understanding of {\it influencer predictions} in evolving social networks; that is, the effects of link prediction on heuristic centrality measures. To the best of our knowledge, link prediction similarity metrics have not yet been used with centrality metrics and $k$ node selection algorithms to approximate the effects of time on a dynamic social network.

We shall seek to quantify influence through simple and complex contagion. As stated by Min \& Miguel \cite{simplecontagion}, simple contagion is a model from epidemiology that runs on the assumption that there is a fixed probability of transmission between each pair of vertices. Centola and Macy \cite{complexcontagiondefinition} defined complex contagion as a model for actions or information with some perceived social cost, considering both personal resistance to influence and forces like peer pressure and social affirmation: a person is infected only if sufficiently many of their neighbors are. 

In general terms, easily accepted information spreads as simple contagion whereas more controversial or effort-requiring topics may be complex contagion. Examples of some categorizations can be seen in \autoref{tab:contagionscenarios} below. We shall begin our study in Section~\ref{sec:background}) by studying link prediction and influencer identification separately on dynamic social networks, and then we shall examine influence through simple and complex contagion models in Section~\ref{subsec:contagionmodels}). By putting the above together we shall present our definition of \textit{\textbf{influencer}}, detail the specific influence model definitions, and introduce our concrete algorithm, the \textbf{Social Sphere Model} (see Section~\ref{sec:ssm}).

\begin{table}[ht]
    \def\arraystretch{1.6}
    \begin{tabular}{ | p{2.5cm} | p{2.5cm}| p{2cm} |}
      \hline
      \textbf{Information Type} & \textbf{Contagion Type} & \textbf{Source} \\
      \hline
      Viral Memes & Simple Contagion & Weng et al. \cite{memescontagions} \\
      Nonviral Memes & Complex Contagion & Weng et al. \cite{memescontagions} \\
      Spread of Disease & Simple Contagion & \\
      Health Behaviors & Complex Contagion & Campbell and Salath'e \cite{healthbehaviors}\\
      Music & Simple Contagion & Notarmuzi et al. \cite{conversationcontroversial} \\
      TV Shows & Simple Contagion & Notarmuzi et al. \cite{conversationcontroversial} \\
      Political/Societal Controversies & Complex Contagion & Notarmuzi et al. \cite{conversationcontroversial} \\
      Social Movements & Complex Contagion & Centola \cite{change}\\
      \hline
    \end{tabular}
    \caption{Example contagion scenarios and categorizations.}
    \label{tab:contagionscenarios}
\end{table}

\begin{figure}[htb]
    \centering
    \begin{subfigure}[b]{0.15\textwidth}
        \centering
        \includegraphics[width=\textwidth]{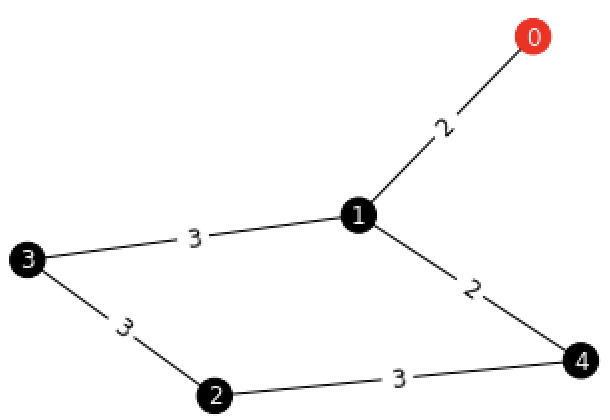}
        \caption{Simple Contagion at time $t=0$}
        \label{fig:sc0}
    \end{subfigure}
    \hfill
    \begin{subfigure}[b]{0.15\textwidth}
        \centering
        \includegraphics[width=\textwidth]{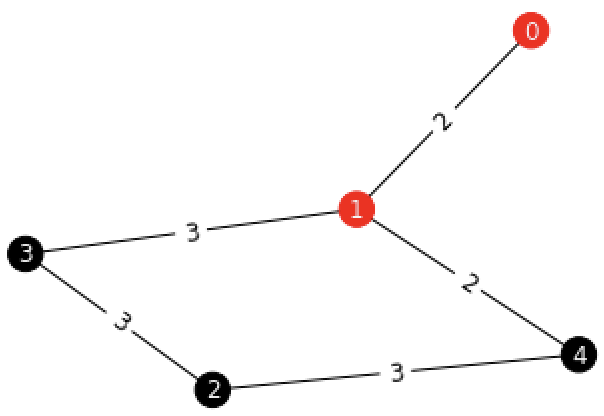}
        \caption{Simple Contagion at time $t=1$}
        \label{fig:sc1}
    \end{subfigure}
    \hfill
    \begin{subfigure}[b]{0.15\textwidth}
        \centering
        \includegraphics[width=\textwidth]{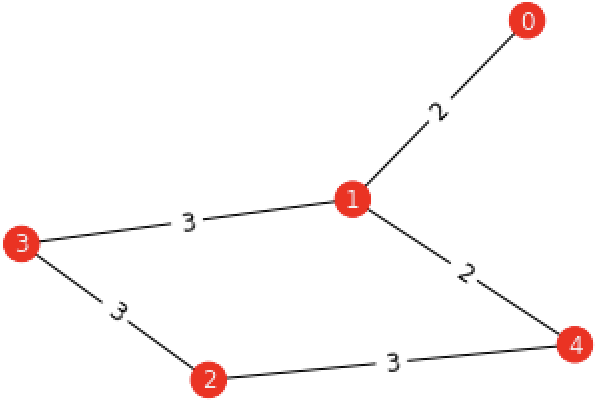}
        \caption{Simple Contagion at time $t=7$}
        \label{fig:sc2}
    \end{subfigure}

    \begin{subfigure}[b]{0.15\textwidth}
        \centering
        \includegraphics[width=\textwidth]{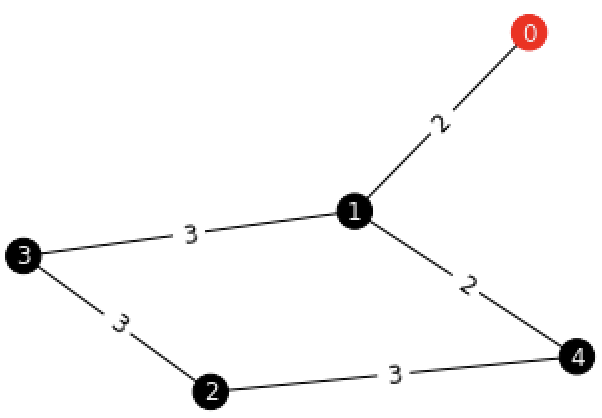}
        \caption{Complex Contagion at time $t=0$}
        \label{fig:cc0}
    \end{subfigure}
    \hfill
    \begin{subfigure}[b]{0.15\textwidth}
        \centering
        \includegraphics[width=\textwidth]{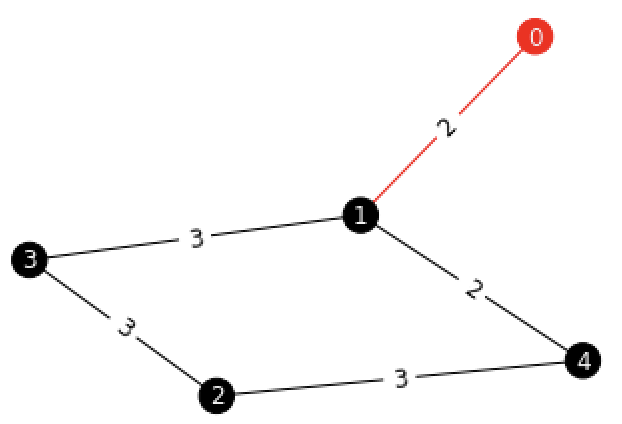}
        \caption{Complex Contagion at time $t=2$}
        \label{fig:cc1}
    \end{subfigure}
    \hfill
    \begin{subfigure}[b]{0.15\textwidth}
        \centering
        \includegraphics[width=\textwidth]{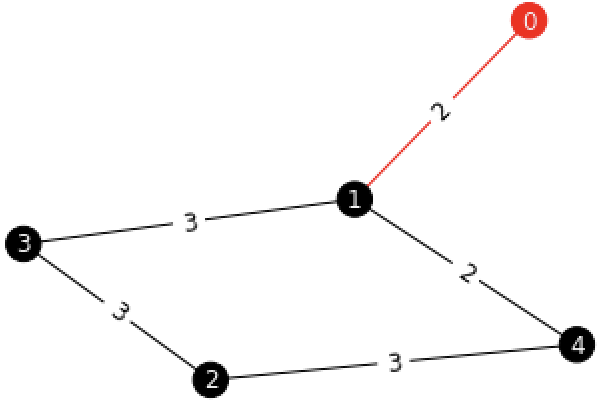}
        \caption{Complex Contagion at time $t=7$}
        \label{fig:cc2}
    \end{subfigure}

    \caption{Contagion examples with initial infected node 0 and distances marked on edges (successful infections and successful interactions for complex contagion are marked with red).}
    \label{fig:contagionex}
\end{figure}

Researchers have noted that the performances of the same set of nodes differ noticeably in the simple and complex contagion (see \autoref{fig:contagionex} for an example). For example, Watts and Dodds \cite{ltmbehavior} found that while the former places a great deal of emphasis on influencers, the latter says away from choosing traditional influencers as initial nodes, especially when considering the relative costs (traditionally influential nodes are likely more expensive hires). 

In Section~\ref{sec:results}, and through the use of  the {\it Social Sphere Model}, we observe that influence generally spreads much faster in the simple contagion model than in the complex. We also notice the following:
\begin{itemize}
    \item better performance for some parameters in our modified prediction metrics and algorithms when compared to the originals (see Section~\ref{sec:background});
    \item high similarity between our model's predicted influencers and the `true' future influencers in both selection and collective influence, suggesting the effectiveness of our model. 
\end{itemize}

Our paper is organized as follows: in Section~\ref{sec:background}, we explore previous works around heuristic measures in link prediction and influencer identification. We explain the Social Sphere Model and our mathematical definition for influencers in Section~\ref{sec:ssm}. Our methodology is detailed in Section~\ref{sec:methods}, with evaluation methods in Section~\ref{subsec:evalmethods}, concrete contagion models in Section~\ref{subsec:contagionmodels}, and experimental setup in Section~\ref{subsec:experimentalsetup}. We examine our model's influence in simple and complex contagion through analysis of our algorithm's results on random graphs in Section~\ref{sec:results}. Finally, we give a summary and future directions for our research in Section~\ref{sec:conclusion}.

\section{Social Networks}\label{sec:background}
Social networks are generally modeled using graph theory by expressing the network as a graph $G = (V,E)$, for    $V$ the set of vertices (the individuals of the network) and $E$ the set of edges (the connections between individuals). Given such a graph, one may assign a weight  function $$w:E\rightarrow [0,1],$$ where $w(u,v)$ is the weight of edge $uv$. Such weights may represent a variety of parameters of the social network, and in this paper, we consider weights to be approximations for probabilities of influence transmission, be it information, disease, or more. In particular, we take inspiration from Goyal et al. \cite{bernoullidistribution}, who took the weights $p(i,j)$ to be the total number of interactions over the total number of trials. In order to study such models, one may assume that the probability of interaction between two nodes during a fixed interval of time is constant, in which case each interval of time can be considered as a Bernoulli trial for temporal datasets.

\subsection{Weight functions} 
Since the geometric distribution counts the number of Bernoulli trials needed to succeed, one can approximate the expected number of intervals needed for an interaction to occur between two nodes $i$ and $j$ as $\frac{1}{w(i,j)}$, the expected value of the geometric distribution with probability $w(i,j)$. These distances are marked on the edges of each graph in \autoref{fig:contagionex}, and they can be used to approximate the amount of time needed for information to spread along each edge. Formally, we can define them as follows. 

\begin{definition}
    The {\it distance} between two vertices $u$ and $v$ though their edge $uv$ is given by the function 
    \begin{eqnarray}d:E &\rightarrow &\mathbb{R}\nonumber \\
uv &\mapsto&   \frac{1}{w(u,v)}.\end{eqnarray}\nonumber
\end{definition}

For convenience, we will also give some preliminary definitions around neighborhoods, paths, and degree:
\begin{definition}
For a graph $G = (V,E)$, define the following:
\begin{itemize}
    \item for $v\in V$, the set of all $n$-order neighbors of $v$ is denoted by $N_n(v)$;
    \item for $u,v \in V$, the set of paths between $u$ and $v$ is denoted by $P(u,v)$;
    \item for a path $$p: (u=v_0, v_1, \dots v_n = v) \in P(u,v),$$ between $u$ and $v$, we define the values 
     \begin{eqnarray}a(p) &=&\sum_{i=0}^{n-1} d(v_i, v_{i+1})\\ b(p) &=& \sum_{i=0}^{n-1} w(v_i, v_{i+1});\\
    D(u,v)& =& \min_{p \in P(u,v)} a(u,v).\end{eqnarray}
\end{itemize}
\end{definition}

To illustrate the above definitions, consider the graphs in \autoref{fig:contagionex}. In this setting, one has that 
 \begin{eqnarray}N_1(0) &=& \{1\}\nonumber\\
 N_2(0) &=& \{3,4\}\nonumber\\P(0,2) &=& \{(0,1,3,2), (0,1,4,2)\}\nonumber\\a((0,1,3,2)) &=& 2+3+3 = 8\nonumber\\b((0,1,3,2)) &=& \frac{1}{2}+\frac{1}{3}+\frac{1}{3} = \frac{7}{6}\nonumber\\D(0,2) &=& 7\nonumber\end{eqnarray} (achieved by the path $p = (0,1,4,2)$). 
 
\begin{remark} The function $D(u,v)$ calculates the minimum weighted distance between vertices $u$ and $v$.\end{remark}

\begin{definition}
    Given a graph $G = (V,E)$, let
    \begin{itemize}
        \item $k_u = |N_1(u)|$ be the unweighted degree of vertex $u$,
        \item $\langle k \rangle = \frac{1}{|V|} \sum_{v\in V} k_v$ be the average degree over all vertices in $G$,
        \item $s_u = \sum_{v \in N(u)} w(u,v)$ be the weighted degree (strength) of vertex $u$, and 
        \item $\langle s \rangle = \frac{1}{|V|} \sum_{v\in V} s_v$ be the average strength over all vertices in $G$.
    \end{itemize}
\end{definition}

Coming back to  \autoref{fig:contagionex}'s underlying graph, for instance, we can see that $$k_0 = 1, ~\langle k \rangle = \frac{1}{5} (1+3+2+2+2) = 2,~  s_0 = \frac{1}{2}$$ 
and $\langle s \rangle = \frac{1}{5} (\frac{1}{2} + \frac{4}{3} + \frac{2}{3} + \frac{2}{3} + \frac{5}{6}) = \frac{4}{5}$.


\subsection{Link Prediction}\label{subsec:linkprediction}
Link prediction is a field that seeks to identify potential future edges in a graph by considering properties of the current graph. There are many types of link prediction strategies, but keeping time complexity in mind we will focus on local methods in this paper, where operating over pairs of nodes will give us similarity scores $s(u,v)$ for all $u\neq v \in V, uv \not \in E$.

The types of link prediction we will be using are {\it path-based}, focusing on first-order and second-order neighbors of nodes. We are interested in how a greater penalty for common neighbors with higher degree can affect performance.

We shall give a brief description of the following link prediction models, which shall allow us in particular to see that they can be thought of as variations or extensions of each other:
\begin{itemize}
    \item common neighbors (e.g. \cite{commonneighbors}),
    \item local path (e.g. \cite{resourceallocation}),
    \item Jaccard coefficient (e.g. \cite{linkpredictioncomparison}),
    \item resource allocation (e.g.  \cite{resourceallocation}),
    \item quasi-local resource allocation (e.g.  \cite{globalextensions}),
    \item RA$-2$, and
    \item quasi-local RA$-2$. \\
\end{itemize}

\noindent \textbf{\textit{Common Neighbors.}} 
Considering collaboration networks, Newman  proposed in \cite{commonneighbors} that the more common neighbors two nodes share, the more likely they are to form a link in the future, leading to the following definition:

\begin{definition}
    The {\it common neighbors (CN) similarity score} for $u,v\in V$ is \[s_{u,v}^{CN} := |N(u) \cap N(v)|.\]
\end{definition}

Murata and Moriyasu \cite{weightedlinkprediction} amended common neighbors for weighted networks to be dependent on the sum of the weights between the nodes summed over common neighbors leading to the following weighted definition: 
\begin{definition}
    The {\it weighted common neighbors (WCN) similarity score} for $u,v\in V$ is \[s_{u,v}^{WCN} := \sum_{x \in N(u) \cap N(v)} \frac{b(u,x,v)}{2}.\]
\end{definition}

\noindent \textbf{\textit{Local Path (Quasi-Local Common Neighbors).}} 
Zhou et al. \cite{resourceallocation} created the local path metric from the observation that common neighbors as a similarity metric often gives the same scores to many nodes, making the actual rankings ambiguous. They extended the computation of score to include paths of three edges, weighting the longer paths by some $\epsilon$:

\begin{definition}
    The {\it  local path (LP) similarity score} for $u,v\in V$ is \[s_{u,v}^{LP} := |N(u) \cap N(v)| + \epsilon |P_2(u,v)|.\]
\end{definition}

Here, we take $\epsilon = 10^{-3}$. Moreover, Aziz et al. \cite{globalextensions} noted that local path is the quasi-local extension of common neighbors.
Extending on the previously mentioned work of Murata and Moriyasu \cite{weightedlinkprediction}, 
Bai et al. \cite{weightedlocalpath} developed a weighted index for local path (weights here are adjusted to match the latter):
\begin{definition}
    The {\it weighted local path (WLP) similarity score } for $u,v\in V$ is \[s_{u,v}^{WLP} := s_{u,v}^{WCN} + \epsilon \sum_{(u,i,j,v) \in P(u,v)} \frac{b(u,i,j) \cdot b(i,j,v)}{4}.\]
\end{definition}

\noindent \textbf{\textit{Jaccard Coefficient.}} 
Liben-Nowell and Kleinberg \cite{linkpredictioncomparison} derived the Jaccard coefficient from information retrieval, with the following definition: 
\begin{definition}
    The {\it Jaccard coefficient (JC) similarity score} for $u,v\in V$ is \[s_{u,v}^{JC} := \frac{N(u) \cap N(v)}{N(u) \cup N(v)}.\]
\end{definition}

Within this setting, a penalty is given when the union of the neighborhoods of the nodes is larger. This can be understood as the more friends people have, the less likely that they will both be with a specific common neighbor and then meet. \\
\\
\noindent \textbf{\textit{Resource Allocation.}} 
Zhou et al. \cite{resourceallocation} came up with the resource allocation metric based on the eponymous process where each individual has resources to be split evenly among its neighbors, defined as follows:

\begin{definition}
    The {\it resource allocation (RA) similarity score} for $u,v\in V$ is \[s_{u,v}^{RA} := \sum_{x \in N(u) \cap N(v)} \frac{1}{|N(x)|}.\]
\end{definition}

Within this model, a penalty is given to the common neighbors with greater degree on the intuition that a shared neighbor with more neighbors itself is less likely to connect those two specific neighbors when it chooses to introduce people on account of having fewer resources to do so.

Lü and Zhou \cite{weightednetworkweakties} made a variant of the above resource allocation for weighted graphs, leading to the following:
\begin{definition}
    The {\it weighted resource allocation (WRA) similarity score} for $u,v\in V$ is \[s_{u,v}^{WRA} := \sum_{x \in N(u) \cap N(v)} \frac{b(u,x,v)}{s(x)}.\]
\end{definition}

\noindent \textbf{\textit{Quasi-Local Resource Allocation.}} 
Aziz et al. \cite{globalextensions} made a quasi-local extension for RA, where one considers paths of three edges in addition to paths of two edges. In other words:
\begin{definition}
    The {\it quasi-local resource allocation (QRA) similarity score} for $u,v\in V$ is {\tiny \[s_{u,v}^{QRA} := \left(\sum_{x \in N(u)\cap N(v)} \frac{1}{|N(x)|}\right) + \epsilon \left(\sum_{(u,i,j,v) \in P(u,v)} \frac{1}{|N(i)||N(j)|}\right).\]}
 
    The {\it weighted quasi-local resource allocation (WQRA) similarity score} for $u,v\in V$ is \[s_{u,v}^{WQRA} := s_{u,v}^{WRA} + \epsilon \sum_{(u,i,j,v) \in P(u,v)} \frac{b(u,i,j)\cdot b(i,j,v)}{s(i)s(j)}.\]\\
\end{definition}

\noindent \textbf{\textit{A New Resource Allocation Variant: RA$-2$.}} 
Finally, returning to the intuition that common neighbors with larger degrees contribute less to the similarity between two nodes, we shall introduce here a new similarity metric as a version of resource allocation. To the best of our knowledge, this is the first time this variant has been introduced. Suppose a node $u$ wants to connect two of its neighbors: there are then $\binom{|N(u)|}{2}$ pairs to choose from, so each pair of its neighbors has a $\frac{1}{\binom{|N(u)|}{2}}$ probability of being linked. For ease of calculation, we approximate this to $$\frac{2}{|N(u)|^2}.$$ Summing over all common neighbors, we then have the RA$-2$ similarity:

\begin{definition}
    The \textit{\textbf{RA$-2$ similarity score}} for vertices  $u,v\in V$ is \[s_{u,v}^{RA2} := \sum_{x \in N(u) \cap N(v)} \frac{2}{|N(x)|^2}.\]
 
    The  \textit{\textbf{weighted RA$-2$ (WRA$-2$) similarity score }} for $u,v\in V$ is  \[s_{u,v}^{WRA2} := \sum_{x \in N(u) \cap N(v)} \frac{b(u,x,v)}{s(x)^2}.\]
\end{definition}

\noindent \textbf{\textit{Quasi-Local variant of RA$-2$}} 
Inspired by Aziz et al. \cite{globalextensions}, we also create a quasi-local variant for RA$-2$:

\begin{definition}
    The \textbf{\textit{quasi-local RA$-2$ (QR$-2$) similarity score}}for $u,v\in V$ is {\tiny \[s_{u,v}^{QR2} := \left(\sum_{x \in N(u) \cap N(v)} \frac{2}{|N(x)|^{2}}\right) + \epsilon \left(\sum_{(u,i,j,v) \in P(u,v)} \frac{4}{|N(i)|^{2}\cdot |N(j)|^{2}}\right).\]}
 
    The \textbf{\textit{weighted quasi-local RA$-2$ (WQR$-2$) similarity score}} for $u,v\in V$ is {\small \[s_{u,v}^{WQR2} := s_{u,v}^{WRA-2} + \epsilon \left(\sum_{(u,i,j,v) \in P(u,v)} \frac{b(u,i,j) \cdot b(i,j,v)}{f(i)^2 f(j)^2} \right).\]}
\end{definition}

In the following sections we shall illustrate the importance of these novel similarity metrics and their use within our Social Sphere Model.

\subsection{Node Centrality}\label{subsec:nodecentrality}

\begin{table*}[t]
    \centering
    \begin{tabular}{ | p{3.5cm} | p{3.5cm} |p{3.5cm} |p{2.5cm}|}
    \hline
    \textbf{Metric} & \textbf{Time Complexity (unweighted)} & \textbf{Time Complexity (weighted)} & \textbf{Type}  \\
    \hline
    Degree Centrality & $O(|E|)$\cite{degreecomplexity} & $O(|E|)$\cite{degreecomplexity} & local \cite{listofinfluencers2} \\
      Coreness & $O(|E|)$\cite{corenessalgorithm} & $O(|E|)$ & iterative \cite{listofinfluencers2}\\
      H-index & $O(|V|\langle k \rangle)$ \cite{h-indexcomplexity} & $O(|V|\langle k \rangle)$ & local \cite{listofinfluencers2} \\
      LocalRank & $O(|V|\langle k \rangle ^2)$ \cite{localrank} & N/A & local \cite{listofinfluencers2} \\
      Clustering Coefficient (and ClusterRank) & $O(|V|\max_{v \in V} k_v^2)$ \cite{clusteringcoefftime} & $O(|V|\max_{v \in V} k_v^2)$ & \cite{listofinfluencers2} \\
      Closeness & $O(|V||E|)$ \cite{timecomplexities} & $O(|V||E| + |V|^2 \log |V|)$ \cite{fibheap} & path-based \cite{listofidentificationtypes}, global \cite{listofinfluencers2} \\
      Betweenness & $O(|V||E|)$ \cite{betweennessalgorithm} & $O(|V||E| + |V|^2 \log |V|)$ \cite{betweennessalgorithm} & path-based \cite{listofidentificationtypes}, global \cite{listofinfluencers2} \\
      Eigenvector & $O(|V|^2\cdot \text{iteration \#})$ \cite{poweriterationmethod} & $O(|V|^2 \cdot \text{iteration \#})$ \cite{poweriterationmethod} & global, iterative \cite{listofinfluencers2} \\
      PageRank & $O(|E| \cdot \text{iteration \#})$ \cite{pagerankcomplexity} & $O(|E| \cdot \text{iteration \#})$ \cite{pagerankcomplexity} & iterative \cite{listofinfluencers2} \\
      LeaderRank & $O(|E| \cdot \text{iteration \#})$ & $O(|E| \cdot \text{iteration \#})$ & iterative \cite{listofinfluencers2} \\
      Balanced Index & $O(\langle k \rangle V + V \log V)$ \cite{ltmmetrics} & $O(\langle k \rangle V + V \log V)$ & local \\
      \hline
    \end{tabular}
    \caption{Time complexities and categorizations of centrality metrics on weighted networks.}
    \label{tab:centralities}
\end{table*}

Vital node identification is a crucial part of social network analysis, applicable to many field ranging from marketing to epidemiology. Centrality methods, some of the most common of these processes, offer a window into understanding influence dynamics within social networks. Mathematically, centrality methods are based  on functions that assign a real-number score to each $v\in V$, essentially ranking the nodes in terms of influence \cite{listofidentificationtypes}. 

The choice of centrality metric often significantly alters the perceptions of who the vital nodes (influencers) are. This section explores common centrality metric definitions, operating principles, and practical implications. We shall give a brief description of the following twelve methods, of which  ~\autoref{tab:centralities} contains a comparative analysis of the time complexities and classifications:
\begin{itemize}
    \item degree;
    \item coreness, introduced by Seidman \cite{kcoreorigin};
    \item H-index, introduced by Hirsch \cite{h-indexorigin};
    \item LocalRank, introduced by Chen et al. \cite{localrank};
    \item ClusterRank, introduced by Chen et al. \cite{clusterrank};
    \item closeness, introduced by Bavelas \cite{closenessorigin};
    \item betweenness, introduced by Freeman \cite{betweennessorigin};
    \item eigenvector, introduced by Landau \cite{eigenvectororigin};
    \item PageRank, introduced by Brin and Page \cite{pagerankorigin};
    \item LeaderRank, introduced by Lü et al. \cite{leaderrank};
    \item balanced index, introduced by Karampourniotis et al. \cite{ltmmetrics}, and
    \item complex path, introduced by Guilbeault and Centola \cite{complexpathlength}.
\end{itemize}

\noindent \textbf{\textit{Degree.}} 
One of the fastest methods to compute node centrality is by considering its graph theoretical degree, creating a centrality metric from the traditional idea that the more connections a vertex has, the more influential it is.
\begin{definition}
    The \textit{degree centrality} of a node $i$ is defined as \[DC_i = k_i.\]
 
    The \textit{weighted degree centrality} (strength) of a node $i$ is defined as \[S_i = \sum_{j \in N(i)} w(i,j) = s_i.\]
\end{definition}

\noindent \textbf{\textit{Coreness.}} 
In 1983, Seidman \cite{kcoreorigin} defined the $k$-core as a measure of the network cohesion of a node, where a subgraph is a $k$-core if it has minimum degree at least $k$. Batagelj and Zaveršnik \cite{corenessalgorithm} define a $k$-core decomposition algorithm in {\bf Algorithm~}\ref{alg:kcore}.

\begin{algorithm}[h]
    \caption{$k$-shell decomposition \cite{corenessalgorithm}}\label{alg:kcore}
    \begin{algorithmic}
    \Require{Graph $G$}
    \State{Sort $V$ by increasing degree}
    \For{each $v \in V$ in order}
        \State{core$[v]$ = deg$[v]$}
        \For{$u \in N(v)$}
            \If{deg$[u] >$ deg$[v]$}
                \State{deg$[u] -= 1$}
                \State{Reorder $u$ in $V$}
            \EndIf
        \EndFor
    \EndFor
    \State{Output list of core values}
    \end{algorithmic}
\end{algorithm}

Later on, Kitsak et al. \cite{coreness} took the core numbers for each vertex as its corenness value. Eidsaa and Almaas \cite{s-core} created a weighted version of $k$-core, where node strength is used rather than degree in the decomposition and when a node $u$ is selected (has the minimum strength in the pruned graph), it is in the $\lceil s_u \rceil$-core. A similar definition can be taken using $\lfloor s_u \rfloor$, and the algorithm is modified accordingly in {\bf Algorithm~}\ref{alg:wkcore} below. \\

\begin{algorithm}
    \caption{weighted $k$-shell decomposition}\label{alg:wkcore}
    \begin{algorithmic}
    \Require{Graph $G$}
    \State{Sort $V$ by increasing strength}
    \For{each $v \in V$ in order}
        \State{core$[v] = \lfloor s(v) \rfloor$}
        \For{$u \in N(v)$}
            \If{deg$[u] >$ deg$[v]$}
                \State{deg$[u] -= w(u,v)$}
                \State{Reorder $u$ in $V$}
            \EndIf
        \EndFor
    \EndFor
    \State{Output list of core values}
    \end{algorithmic}
\end{algorithm}

\noindent \textbf{\textit{H-index.}} 
Along the same lines of study, but within the academic world, Hirsch \cite{h-indexorigin} proposed the H-index to measure the impact of a scientist's research outputs through considering citations per paper. Lü et al. \cite{h-index} then used it in the context of centrality metrics, introducing the following definition:

\begin{definition}
Define the operator $H$ on a finite set $S = \{x_1, \dots , x_m\}$ as the maximum integer $h$ such that there are $h$ elements in $S$ with values $\ge h$. The {\it $n$-order H-index} is defined as \[h_i^{(n)} = H(h_{j_1}^{(n-1)}, \dots , h_{j_{k_i}}^{(n-1)}),\] where $h_i^{(0)} = DC_i$.
\end{definition}

From the above perspective, the classical H-index can be seen as the $1$-order H-index. Lü et al. \cite{h-index} proved that the H-index forms a middle ground between a progression from degree centrality to coreness, where the $0$-order index is degree and the $\infty$-order index converges to the $k$-core score. Finally, it is interesting to note that Zhao et al. \cite{weightedhindex} defined the w-lobby index, a metric for weighted networks similar to the H-index, as follows:
\begin{definition}
    The \textit{w-lobby index} of a node $u$ is the largest integer $k$ such that $|\{s(v) \ge k | v \in N(u)\}| \ge k$.
\end{definition}

Through the above definitions, we can introduce the \textit{$n$-order H-degree}, which is defined as \[WH_i^{(n)} = WH(wh_{j_1}^{(n-1)}, \dots , wh_{j_{k_i}}^{(n-1)}),\] with $wh_i^{(0)} = s(i)$. In this paper, we take the $10$th order H-index.

\noindent \textbf{\textit{LocalRank.}} 
Aiming to find a more effective local method due to the high time complexities of global and path-based methods, Chen et al. \cite{localrank} introduced the following extension of degree centrality to more than just first-order neighborhoods:
\begin{definition}
    The \textit{LocalRank} score of a node $u$ is \[LR_u = \sum_{v \in N(u)} \sum_{w \in N(v)} |N(w)| + |N_2(w)|.\]
\end{definition}

\noindent \textbf{\textit{ClusterRank.}} 
Watts and Strogatz \cite{clusteringcoeff} defined the clustering coefficient around the percent of existing links between a node's neighbors over the total potential amount to quantify how clustered the network is:
\begin{definition}
    The {\it clustering coefficient} of a node $i$ in the directed network $c_i$ is defined as \[c_i = \frac{|\{(jk \in E |j, k \in N(i)\}|}{k_i (k_i-1)}.\]
\end{definition}

Working with the intuition that nodes in more clustered parts of the network may be more redundant and thus have less influence, Chen et al. \cite{clusterrank} then used the clustering coefficients of nodes to define the following centrality metric:

\begin{definition}
    The \textit{ClusterRank} score of a node $i$ is \[f(c_i) \sum_{j \in N(i)} (k_j + 1),\] where $f(c_i)$ is an exponential function of $c_i$ of the form $\alpha^{-c_i}$ for a fixed constant $\alpha$ (usually $10$).
\end{definition}

\begin{remark}This metric considers the clustering coefficient of the neighboring parts of the network: a node whose neighbors are already tightly clustered may be able to spread information within its cluster very quickly, but has more difficulty doing outreach to other communities, whereas nodes that may bridge clusters might be better.\end{remark}

Taking the weighted score to be $f(c_i) \sum_{j \in N(i)} (s_j + 1)$ gives a modification for weighted networks.\\
 
\noindent \textbf{\textit{Closeness.}} 
Bavelas introduced  in 1948 \cite{closenessorigin} the idea of closeness centrality as 
proportional to the reciprocal of the sum of the shortest path lengths from each pair of vertices $u\neq v$. Rephrasing, Freeman \cite{closeness} defines the concept of closeness centrality as a measure of the relative proximity of a node $p_k$ to the rest of the graph $G$, defined as $$C_C(p_k) = \frac{n-1}{\sum_{i=1}^n D(p_k,p_i)},$$ where $V = \{p_1,\dots , p_n\}$ are the vertices of the graph. 

In a survey of influencer identification metrics, Lü et al. \cite{listofidentificationtypes} note that disconnected networks then pose a problem as distances go to infinity, making the nodes indistinguishable by closeness centrality (as they are all then $0$). They then give the following modification:
\begin{definition}
    The \textit{closeness centrality} of node $i$ is defined as the reciprocal of the harmonic mean of the geodesic distances from $i$ to all other nodes, i.e. $$CC_i = \frac{1}{n-1}\sum_{j\neq i}\frac{1}{d_{i,j}}.$$
\end{definition}

\noindent \textbf{\textit{Betweenness.}} 
Betweenness centrality considers how critical a node $i$ is to the spread of information, e.g. what percentage of shortest paths going from one node to another must pass through $i$. First formally defined it in 1977 by Freeman \cite{betweennessorigin}, it can be understood along with an efficient algorithm for computing betweenness centrality, through the work of Brandes \cite{betweennessalgorithm}:
\begin{definition}
    The \textit{betweennness centrality} of a node $i$ is simply $$BC_i = \sum_{s,t \neq i, s\neq t} \frac{g_{st}^i}{g_{st}},$$ where $g_{st}^i$ is the number of paths that pass through $i$ among the shortest paths between $s$ and $t$ and $g_{st}$ is the number of shortest paths between $i$ and $j$.
\end{definition}

\noindent \textbf{\textit{Eigenvector.}} 
In \cite{eigenvectororigin} Landau  gave a clarified overview of a previous paper of his detailing the use of eigenvectors to rank chess players, which has been credited with being the first use of eigenvector centrality. Later on,
Lü et al. \cite{listofidentificationtypes} defined the eigenvector centrality scores $EC$ of $G$ as the vector $\vec{x}$ with components $x_i=EC_i$ such that $\vec{x} := \frac{1}{\lambda} A \vec{x}$ for the largest eigenvector $\lambda$ of $A$. In other words,

\begin{definition}
    The \textit{eigenvector centrality} of a node $i$ is the $i$th component of the eigenvector corresponding to the largest eigenvalue of $A$.
\end{definition}

The assumption behind this is that a node's influence is a linear combination of the influences of its neighbors, weighted by the weights of its outgoing edges.\\
\\
\noindent \textbf{\textit{PageRank.}} 
PageRank is a search algorithm presented by Brin and Page \cite{pagerankorigin} in a paper where they introduced a Google prototype, modeled on the behavior of a bored, link-clicking user. They defined the algorithm as follows:
\begin{definition}
    The PageRank score $PR(u)$ of each vertex $u$ satisfies the following property:
    \[PR(u) = (1-d) + d\sum_{v\in N(u)} \frac{PR(v)}{k_v},\] where $d$ is a fixed constant in $[0,1]$.
\end{definition}

They set $d$ as the damping coefficient, where the $(1-d)$ term represents the probability a user loses interest and goes to a random webpage. For our purposes, we will set $d = 1$.

Xing and Ghorbani \cite{weightedpagerank} modified the PageRank algorithm for weighted graphs: \\
\begin{definition} \label{def:wpr}
    The \textit{weighted PageRank} score $PR(u)$ of each vertex $u$ satisfies the following property:
    \[WPR(u) = (1-d) + d\sum_{v\in N(u)} \frac{WPR(v) \cdot w(u,v)}{s_v},\] where $d$ is a fixed constant in $[0,1]$.
\end{definition}

\noindent \textbf{\textit{LeaderRank.}} 
Inspired by PageRank, Lü et al. \cite{leaderrank} created the LeaderRank algorithm. Their intention was to include a ``ground node" connected to every node in the network, called node $n+1$, and then used the following recursive relationship to solve for their influences:
\begin{definition}
    At time $t$, the \textit{LeaderRank} score of node $i$ is defined recursively as $LR_v(t) = \sum_{v \in N(u)} \frac{LR_v(t-1)}{k_v}$.
\end{definition}

To differentiate the original nodes from the added one, all nodes from the original graph begin with value $1$ and the leader node begins with value $0$. Once the function stabilizes, the leader node's value gets evenly distributed to the rest of the nodes and deleted. Thus, the LeaderRank scores at the ending time $t_c$ are $LR_u = LR_u(t_c) + \frac{LR_{n+1}(t_c)}{|V|}$ for each vertex $u \in V$. Weighted LeaderRank scores can be calculated in the same fashion as weighted PageRank (see Definition~\ref{def:wpr}). \\
\\
\noindent \textbf{\textit{Balanced Index (BI)}} \\
Karampourniotis et al. \cite{ltmmetrics} introduced the Balanced Index metric for complex contagion by taking a weighted combination the metrics of resistance, degree, and neighbors' degrees. In the context of threshold models, the resistance $r_v$ of a node $v$ is defined as a measure of the current threshold of each node, where uninfected nodes $v$ initially begin with $r_v = \theta_v \cdot s(v)$ and experience decreasing resistance as more of its neighbors become infected. At resistance $0$ the node also becomes infected. The neighbors' degree comes from Karampourniotis et al.'s \cite{ltmmetrics} indirect drop of resistance metric, which for node $v$ takes the sum over the nodes that $v$ could infect in a single turn (i.e. the nodes $j$ with remaining resistance less than $\frac{w(i,j)}{s(i)}$). The balanced index metric is thus as follows:

\begin{definition}
    The \textit{balanced index} score of a node $i$ is \[BI_i = a\cdot r_i + b\cdot k_i + c \sum_{j \in N(i) | r_j = 1} (k_j - 1)\] for nonnegative weights $a+b+c = 1$.
\end{definition}

We can adjust this for weighted graphs by taking \[BI_i = a\cdot r_i + b\cdot s_i + c \sum_{j \in N(i) | r_j \le w(i,j)} (s_j - w(i,j)).\]
\\
\noindent \textbf{\textit{Complex Path Centrality.}} 
Seeking to extend the path length measures traditionally used in simple contagion to the wide bridges (reinforcing ties) used in complex contagion paths, Guilbeault and Centola \cite{complexpathlength} created complex path centrality, based on the following:

\begin{definition} Let $T_j$ be the resistance of node $j$. Then, 
 \begin{itemize}
    \item the bridge between nodes $i$ and $j$ (the set of nodes in $N(j)$ that are distance $0$ or $1$ away from some node in the neighborhood of $N(i)$) is 
    \[BW_{i,j} := \{v | v \in N(j) \land (v \in N(i) \lor \exists u \in N(i) : uv \in E)\},\]
with width $W_{i,j} = |B_{i,j}|$ and indicator variable $[W_{i,j}]$ such that $[W_{i,j}]=1$ if $W_{i,j} \ge T_j$ and $0$ otherwise. A bridge is locally sufficient for contagion spread if $W_{i,j} \ge T_j$.
    \item $B_i$ is the set of nodes $j\neq i$ such that $W_{i,j} > 0$,  
    \item the proportion of locally sufficient bridges is $LB_i = \sum_{x\in B_i} [W_{i,j}]$,
    \item $GEO_{i,j}$ is the shortest path between $i$ and $j$ only through edges between nodes connected by sufficient bridges, and 
    \item $\phi(GEO_{i,j})$ is the list of vertices in the path $GEO_{i,j}$.
\end{itemize}
\end{definition}

From this, they then defined a complex path from node $i$ to node $j$ as the sequence of sufficiently wide paths from $i$ to $j$. Mathematically, they write it as follows:
\begin{definition}
The \textit{complex path centrality} of a node $i$ is defined as \[CC_i = \frac{1}{n - k_i}\sum_{i\neq j} |\phi(GEO_{i,j})|.\]
\end{definition}

\subsection{Heuristic Top-$k$ Algorithms}\label{subsec:algorithms}
In marketing, one generally finds it prudent to advertise through more than one channel; thus we are concerned with not only finding the single most effective influencer, which centrality metrics aim to do, but with finding the set of $k$ influencers with the greatest collective influence for some $k$ determined by budget.

Transitioning from the single influencer concept to this more practical problem then introduces more complexities. In particular, from Lü et al. \cite{listofidentificationtypes}, the straightforward approach of taking the highest $k$ nodes by score in a given centrality metric could lead to redundant influences. When information travels relatively unresisted, as in simple contagion (see  \autoref{tab:contagionscenarios}), redundancy is often undesirable, as when a single successful interaction suffices to influence an individual, having too much overlap between the chosen nodes' targets potentially undermines their collective effectiveness and influence. However, in complex contagion, social reinforcement is actually necessary to induce influence cascades, which leads to interesting questions around the straightforward top-$k$ algorithm. We shall thus assess whether the perceived redundancy truly implies reduced performance.

In this section, we explore alternative methodologies to reduce redundancy in the locations of the chosen influencers and modify them to consider centrality metrics as well. We consider the following five algorithms and modify some of them to incorporate centrality metrics, as different centrality metrics may have underlying properties that perform better in tandem with certain algorithms. For the sake of convenience, call these modified algorithms as \textit{\textbf{centrality algorithms}}. \smallbreak
 
\noindent \textbf{\textit{LIR.}}
To reduce the number of adjacent nodes in the chosen set of $k$, Liu et al. \cite{lir_original} created the local index rank (LIR) algorithm. They first defined the function $Q(x)$ to be $1$ for $x > 0$ and $0$ otherwise, then set the $LI$ score of each node $u$ as \[LI(u) = \sum_{v\in N(u)} Q(k_v-k_u). \]
They then focused on the nodes with $LI$ value $0$ (disregarding those that simply have no edges) and picked the $k$ nodes from there with largest degree \footnote{In this paper, we modify the original algorithm to incorporate centrality metrics by sorting by the score the node has on a given centrality metric.}.\\

\begin{algorithm}
\caption{LIR Algorithm \cite{lir_original}} \label{alg:lir_og}
    \begin{algorithmic}
        \Require{$G$, $k$}
        \State{Find the $LI$ values for all the nodes}
        \State{Find the list $\{v | LI(v) = 0\}$}
        \State{Sort nodes by degree (decreasing)}
        \State{Output the top $k$ nodes in the sorted list}
    \end{algorithmic}
\end{algorithm}

\noindent \textbf{\textit{LIR-2.}}
Inspired by the LIR method described before, Tao et al. \cite{lir2} created an LIR-2 method as a quasi-local extension to second-order neighbors. They define the LIR-2 score of a node $u$ to be $$LI_2(u) = \sum_{v\in N(u)} Q(k_v-k_u),$$ as shown in {\bf Algorithm~}\ref{alg:lir_2}.

\begin{algorithm}
\caption{LIR-2 Algorithm \cite{lir2}} \label{alg:lir_2}
    \begin{algorithmic}
        \Require{$G$, $k$}
        \State{Find the $LI_2$ values for all the nodes}
        \State{Make sorted list $L$ of sets of nodes with the same $LI_2$ values in increasing order}
        \State{Make an empty list FinalList}
        \For{Set $S$ in $L$}
            \State{Sort nodes in $S$ in decreasing order of degree}
            \State{Append to FinalList}
        \EndFor
        \State{Output the top $k$ nodes in FinalList}
    \end{algorithmic}
\end{algorithm}

\noindent \textbf{\textit{Joint Nomination.}}
Dong et al. \cite{jointnomination} proposed the joint nomination algorithm as an algorithm geared towards immunization purposes, where it seeks to select vertices whose removal can help stop influence spread. Their method spans $k$ rounds. In each round, a random node is selected as a nominator, one of its neighbors is randomly selected as conominator, and finally a random common neighbors of the nominator and conominator is selected into the chosen set, as shown in {\bf Algorithm~}\ref{alg:jointnom}. For developing our  \textit{\textbf{Social Sphere model}} we shall amend the method of random selection by including probabilities based on centrality metric score and weights \footnote{In this paper, we deviate from the original version of joint nomination by choosing nodes with probability proportional to centrality score, choosing co-nominators with probability proportional to the weight of the edges between them, and choosing a common neighbor (nominee) $i$ with probability $\frac{w(u,i) \cdot w(v,i)}{\sum_{j\in N(u) \cap N(v)} w(u,j)\cdot w(v,j)}$.}.\smallbreak

\begin{algorithm}
\caption{Joint Nomination \cite{jointnomination}} \label{alg:jointnom}
    \begin{algorithmic}
    \Require{$G, k$}
    \State{Create a set of nominees}
    \State{Identify a set $N$ of $k$ random nominators}
    \For{$u \in N$}
        \State{Identify set of neighbors $C$}
        \State{Randomly pick a co-nominator $v$, where neighbor $i$ is chosen with weight $\frac{w(u,i)}{s(u)}$}.
        \While{$N(u) \cap N(v)$ is $\emptyset$}
            \State{Remove $v$ from $C$}
            \State{Randomly pick another co-nominator $v$}
            \If{$C$ is empty}
                \State{Add a random conominator to the nominees}
                \State{Break}
            \EndIf
        \EndWhile
        \If{$|N(u) \cap N(v)| > 0$}
            \State{Randomly pick a nominee $n$ from their common neighbors}
            \State{Add $n$ to the nominees}
        \EndIf
    \EndFor
    \State{Output nominees}
    \end{algorithmic}
\end{algorithm}
\noindent \textbf{\textit{VoteRank.}} 
From a different perspective, Zhang et al. \cite{voterank} proposed VoteRank as an algorithm where influential nodes are chosen by ``votes" of the vertices in the graph. There are $k$ rounds of voting. Each node begins with voting power $1$ and score $0$ and will divide their voting power among their neighbors, thus changing their scores. In this setting, the node with maximum score is chosen in each round and a resulting penalty given to its neighbors (see {\bf Algorithm~}\ref{alg:voterank}).

\begin{algorithm}
\caption{VoteRank Algorithm \cite{voterank}} \label{alg:voterank}
    \begin{algorithmic}
        \Require{$G$, $k$}
        \State{Initialize each node $v$ with score $score(v) = 0$ and voting ability $va(v) = 1$}
        \State{Create an (empty) list $L$ of chosen nodes}
        \For{$i$ in range$(k)$}
            \State{Reset the scores of all the nodes to $0$}
            \For{$u \in V \setminus L$}
                \For{$v \in N(u) \setminus L$}
                    \State{Add $1$ to $score(v)$}
                \EndFor
            \EndFor
            \State{Pick the node $n$ with greatest score and add it to $L$}
            \State{Set the voting power of $n$ to $0$}
            \State{Decrease the voting power of its neighbors by $\frac{1}{\langle s \rangle}$}
        \EndFor
        \State{Output $L$}
    \end{algorithmic}
\end{algorithm}

Later on, researchers such as Sun et al. \cite{wvoterank} have modified VoteRank to fit weighted graphs by changing the score to include the importance of having many neighbors. Researchers such as Li et al. \cite{dilvoterank} and Kumar and Panda \cite{kcorevoterank} have combined VoteRank with metrics such as the DIL method (proposed by Liu et al. \cite{dil}) and coreness; inspired by these works, we give a generalization of VoteRank for a given centrality metric $c$ (such as the ones in Section~\ref{subsec:nodecentrality}) by setting each node's voting ability to its score with respect to $c$.\\

\noindent \textbf{\textit{Graph Coloring.}} 
Finally, aiming to avoid redundancy, Zhao et al. \cite{graphcoloring} proposed the graph coloring method as a means of selecting separated vertices. They use the Welsh-Powell algorithm to separate the graph into sets such that each set has a unique color and no nodes in the same set share an edge. For a given centrality metric, they then pick the $k$ nodes with highest score from the largest independent set as their chosen $k$ vertices. 

\begin{algorithm}
\caption{Welsh-Powell Algorithm \cite{graphcoloring}}\label{alg:welshpowell}
\begin{algorithmic}
\Require{Graph $G$}
\State{Label nodes as $v_1, \dots , v_n$ in descending order according to degree}
\State{Let $\pi (v_1) = 1$}
\For{$i$ in $\{1,2,\dots , n-1\}$}
    \State{Let $C(v_{i+1}) = \{\pi(v_j) | j\le i, v_j \in N(v_{i+1})\}$}
    \State{Take the smallest numbered color in the set of colors $C$ not in $C(v_{i+1})$ as $\pi(v_{i+1})$}
    \State{$i+=1$}
\EndFor
\State{Output the values of the color function $\pi$}
\end{algorithmic}
\end{algorithm}
Through the extension and modification of the above algorithms, we shall proceed to introduce our \textbf{Social Sphere Model}. 

\subsection{Comparison to GNN-Based and Temporal Deep Learning Approaches}

Recent advances in link prediction and influence maximization have made significant use of geometric deep learning (GDL), particularly Graph Neural Networks (GNNs). These models learn node embeddings that capture complex structural and temporal features of evolving networks. GNN-based models such as GCNs, GraphSAGE, and temporal variants like DyGCN, TGAT, and TGN have shown strong performance in tasks like dynamic link prediction and influence estimation \cite{info13030123, NEURIPS2018_53f0d7c5, hamilton2020grl}.

However, these approaches come at a cost as they typically require extensive training on large datasets and GPU resources; their internal decision processes can be difficult to interpret, limiting transparency; and they often require temporal labels or timestamps that may not be available. In contrast, the Social Sphere Model relies on interpretable, path-based link prediction scores and top-k heuristics to identify future influencers. While it may not match GNNs in raw predictive power on large datasets, its strengths lie in:
\begin{itemize}
    \item Reduced computational complexity;
    \item Transparency and explainability of results;
    \item Applicability to small-scale or under-sampled temporal networks.
\end{itemize}

This trade-off between interpretability and model expressiveness has been acknowledged in the literature \cite{rep_learning_graphs}, with ongoing research attempting to bridge the gap.

In future work, we aim to extend our framework by integrating learned embeddings into the prediction stage or applying our interpretability heuristics as post hoc analyses to GNN-based results.

\section{The Social Sphere Model} \label{sec:ssm}
Expanding on the previous sections, we shall introduce here our model, the \textbf{\textit{Social Sphere Model}} for path-based similarity and heuristic centrality measures, an approach designed here to predict future influencers in changing social networks. This model synthesizes key elements of link prediction and top-$k$ influencer algorithms, offering a comprehensive tool for understanding and anticipating changes in social networks. The uniqueness of our approach lies in the use of lower complexity link prediction metrics rather than more costly neural networks and the inclusion of expected value in our predicted graph creation.\\

\subsection{Mathematical Influencers} 
As mentioned before, one may study social networks through their graph representation, where vertices represent individuals, and their relations are reflected within the edges. {In influence maximization, influencers are often defined as those most important to a network, whether structurally or influentially \cite{survey2}. In this paper, we shall use a more specific definition of influencer in relation to centrality and influence maximization algorithms.} \\
\begin{definition}[Influencer]\label{influencer}
    A vertex in a network represented by a graph $G$ is an influencer (with respect to some centrality metric) if it is either a $k$-influencer or a single influencer, where
    \begin{itemize}
        \item a \textit{$k$-influencer} is a vertex selected through one of the $k$-node selection algorithms with respect to some centrality metric, and
        \item a \textit{single influencer} is the highest scoring vertex with respect to some centrality metric.
    \end{itemize}
\end{definition}

\subsection{Future Network Prediction} 
Understanding how to predict the future state of influencers is of upmost importance when considering information spread. Through previously reviewed literature, one is able to do the following:  
\begin{itemize}
\item predict future links of our social network, and 
\item find the most influential $k$ nodes of a given social network. 
\end{itemize}
We shall combine these and look for future influencers by first predicting the state of the future network, like Yanchenko et al. \cite{exante} did. However, when considering the formation of the future graph, rather than selecting the top edge pairs, we simply take normalized similarity scores to be probabilities of transmission and add them to the graph as weights, using expected value to calculate the distance. Moreover, given the unit of time $t$ we are predicting for, we can find a better predictor for the transmission probability.

Suppose the probability an edge forms between two nodes $u$ and $v$ in one unit of time is $p_{u,v}$. Consider the analogy of a coin flip: an edge forming between two nodes in one unit of time is equivalent to a correspondingly weighted coin landing heads on a single flip. Thus, the probability an interaction has occurred in those $t$ units of time is simply the probability the coin comes up heads some time in its $t$ flips, which is $1-(1-p_{u,v})^t$. Similarly, we can take the outcome of the coin flip as whether or not two nodes become connected. Then the probability of connection, and thus transmission of information, is simply $1-(1-p_{u,v})^t$, which we take as an updated weight, enabling us to better account for time evolution.  
 
\subsection{Future Top-$k$ Nodes Prediction} 
Our algorithm for future influencer prediction (see {\bf Algorithm~}\ref{alg:futureinfluencers}) takes a graph and, given a target $t$ units of time in the future, predicts future connections with a normalized similarity metric. It then applies a top-$k$ algorithm to identify potential future influencers. \\

\begin{algorithm}
\caption{Future Top-$k$ Nodes Prediction} \label{alg:futureinfluencers}
    \begin{algorithmic}
        \Require{Graph $G$, Normalized Similarity Metric $M$, Top-$k$ Algorithm $c$, $t$}
        \State{For vertices $u \neq v$ with $uv \not \in E$, calculate $s^M_{u,v}$.}
        \For{$u,v$ such that $s^M_{u,v} \neq 0$}
            \State{Add edge $uv$ to $G$ with weight $1-(1-s^M_{u,v})^t$ and distance $\frac{1}{1-(1-s^M_{u,v})^t}$.}
        \EndFor
        \State{Identify the top $k$ nodes on $G$ using $c$.}
    \end{algorithmic}
\end{algorithm}

In what follows we shall first described the methods developed in order to implement our model, give an analysis of our results, and describe some of the applications and future directions.  

\section{Methods}\label{sec:methods}
In order to illustrate the different uses of our model and our algorithm, we shall test our algorithm on a random graph: given a graph $G$, we form a training graph from a random subset of existing edges of given size, predict the future state of the training graph, and identify influencers on those predicted graphs. For each graph and parameter tested, we repeated the process ten times, averaging the data. We modelled a dynamic network by taking the training graph as the current configuration and the original graph $G$ as the future network, thus allowing us to assess the efficacy of the influencers found in terms of their ability to ``infect'' nodes over time when set as the initial spreaders in $G$. 

\subsection{Evaluation Methods} \label{subsec:evalmethods}
Our testing focused on two key aspects: \begin{itemize}\item[(I)] the effectiveness of the predicted influencers compared to those found in the actual network, and\item[(II)] the success of our predictions in identifying true influencers.\end{itemize}

To measure the effectiveness mentioned in \textbf{(I)}, we quantify influence as the percentage of nodes infected at each unit of time when the chosen influencers are set as initial spreaders on $G$ and an influence model (see Section~\ref{subsec:contagionmodels}) is run. The results are then presented visually in the form of line graphs. From a different perspective, we describe the predictive success mentioned in \textbf{(II)} as the similarity between our predicted influencers (and their influence) and the influencers found on $G$ (the ``true" influencers), quantified in two different ways:
\begin{enumerate}
    \item {\it accuracy}, the fraction of chosen influencers on the predicted graph that are also chosen by the same methods of influencer identification on the original, and
    \item {\it mean squared error}, a function from statistics that evaluates the differences between the influence of the influencers chosen on the predicted graphs and the original graph; for a given prediction metric and influencer identification algorithm, the mean squared error is \[\frac{1}{r} \sum_{i=0}^r (O(t) - P(t))^2,\] where $P(t)$ is the fraction of infected nodes at time $t$ when the influence of the influencers chosen on the predicted graph are evaluated on the original graph and $O(t)$ is defined similarly for the original graph.
\end{enumerate}

\subsection{Contagion Models}\label{subsec:contagionmodels}
To account for different types of information, we evaluate influence according to two simplified models for {\it simple contagion} and {\it complex contagion}. \\

\noindent \textbf{Simple Contagion. }
In general, simple contagion can be understood as an ``infection" of information, where an uninfected individual $j$ becomes infected with probability $p$ once news is passed along to them from an infected individual $i$, in a simple percolation model. For simplicity's sake, we shall take $p=1$ (following SRI models). Then consider our definition of distance as the expected units of time transmission takes; with this model, influence thus can be approximated as transmitting across each edge $uv$ in $d(u,v)$ units of time. Thus, we can see the following: 

\begin{definition}
 In a {\it Simple Contagion} model, at time $t$, the infected set $I(t)$ is defined as follows:
    \[I(t) = \{v\in V | \exists u \in I(0) : D(u,v) \le t \}.\]
\end{definition}

\noindent \textbf{Complex Contagion.}
Complex contagion is usually described by threshold models, where each node is given a resistance $\theta_v$ and is activated according to a function \[f:I \rightarrow [0,1]\] (see e.g. Shakarian et al. \cite{icmltm} for further details). While resistance often varies depending on the individual, in this paper we set it to be a fixed percentage of each vertex's strength, as considering how perceived plurality can lead to adoption, individuals with larger neighborhoods may have more resistance and similarly the converse may happen. 

\begin{definition}
  In a {\it Complex Contagion} model, at time $t$, the infected set $I(t)$ is defined as follows:
    \[I(t) = \{v\in V | \sum_{i\in I(t-1)} w(i,v) \ge \theta \cdot s_v\},\]
    where $\theta \in [0,1]$ is a given threshold fraction.
\end{definition}

\subsection{Experimental Setup}\label{subsec:experimentalsetup}
We tested our algorithm on an Erdös-Rényi random graph and an existing collaboration network dataset. Erdös and Rényi \cite{erdosrenyirgraph} defined these random graphs by taking two parameters, $n\in \mathbb{Z}$ and $p \in [0,1]$. The graph is created with $n$ vertices, and each pair of vertices is connected by an edge with probability $p$. We used a graph with $n=500$ and $p = 0.05$. In the random graph case, we took $k=5$. Additionally, we tested the influences of single influencers in each scenario as well.

For each of the datasets, we took two parameters: one when $90\%$ of edges are chosen and $t=1$, and the second one when $70\%$ of the edges are chosen and $t = 3$. We are assuming that edges occur at around the same rate each unit of time; thus, the former models influencer prediction in the near future and the latter predicts for more distant networks.

This methodology and experimental setup pave the way for our findings, which offer novel insights into the prediction of influencers in dynamic social networks.

\section{Results}\label{sec:results}
Thorough applications of our model to  Erdö-Rényi random graphs, one can see the following:
\begin{itemize}
    \item the average MSE for each of our algorithms is low, indicating a high similarity to the performance of influencers chosen on the ``future" graph,
    \item the average accuracy differed depending on algorithm and Social Sphere Model parameters, but is often quite high, and
    \item the influence of influencers from the predicted graphs is comparable to that of the original and often performes better than influencer identification done on their training graphs.
\end{itemize}

\begin{table}[htp]
\centering
    \begin{tabular}{lccc}
    \hline
    \textbf{Prediction Metric} & \textbf{Complex} & \textbf{Simple} & \textbf{Overall} \\
    \hline
    Common Neighbors & 0.10224 & 9.77175E-05 & 0.05117 \\
    Jaccard & 0.10427 & 0.00011 & 0.05219 \\
    Local Path & \textbf{0.09998} & 0.00020 & \textbf{0.05009} \\
    Quasi-Local RA & 0.11917 & 0.00019 & 0.05968 \\
    Quasi-Local RA$-2$ & 0.10407 & 0.00020 & 0.052134 \\
    RA$-2$ & 0.10120 & 0.00010 & 0.05065 \\
    Resource Allocation & 0.12174 & \textbf{9.47277E-05} & 0.06092 \\
    \hline
    Overall & 0.10753 & 0.00014 & 0.05383 \\
    \hline
    \end{tabular}
    \caption{Average MSE over all top-$k$ algorithms for link prediction metrics using $70\%$ training graphs from a $500$-node Erdös-Rényi graph.}
    \label{tab:rg_sms_mse_70}
\end{table}

\begin{table}[htp]
\centering
    \begin{tabular}{lccc}
    \hline
    \textbf{Prediction Metric} & \textbf{Complex} & \textbf{Simple} & \textbf{Overall} \\
    \hline
    Common Neighbors & 0.10127 & \textbf{7.30107E-05} & 0.05067 \\
    Jaccard & 0.09809 & 7.47762E-05 & 0.04908 \\
    Local Path & 0.07243 & 0.00020 & 0.03632 \\
    Quasi-Local RA & 0.07870 & 0.00020 & 0.03945 \\
    Quasi-Local RA-2 & \textbf{0.07062} & 0.00020 & \textbf{0.03541} \\
    RA-2 & 0.10448 & 7.62052E-05 & 0.05228 \\
    Resource Allocation & 0.10244 & 7.05446E-05 & 0.05126 \\
    \hline
    Overall & 0.08972 & 0.00013 & 0.04492 \\
    \hline
    \end{tabular}
    \caption{Average MSE over all top-$k$ algorithms for link prediction metrics using $90\%$ training graphs from a $500$-node Erdös-Rényi graph.}
    \label{tab:rg_sms_mse_90}
\end{table}

\begin{table*}[t]
    \centering
    \begin{tabular}{ | p{2.5cm} | p{1.7cm} |p{1.5cm} |p{1.2cm} |p{1.2cm} |p{1.2cm} |p{1.2cm} |p{1.7cm} |p{1.2cm} |}
    \hline
    Algorithm & Common Neighbors & Jaccard & Local Path & Quasi-Local RA & Quasi-Local RA-2 & RA-2 & Resource Allocation & Overall \\
    \hline
    Centrality VoteRank & 0.32 & 0.3467 & 0.3183 & \textbf{0.3567} & 0.345 & 0.345 & 0.3533 & 0.3407 \\
    Graph Coloring & \textbf{0.0167} & \textbf{0.0167} & 0.01 & 0.01 & 0.01 & \textbf{0.0167} & \textbf{0.0167} & 0.01381 \\
    Joint Nomination & 0.0117 & 0.0183 & 0.0117 & 0.0183 & 0.0133 & \textbf{0.025} & 0.0117 & 0.0157 \\
    k Highest & 0.4567 & 0.4617 & 0.4183 & 0.4267 & 0.4417 & \textbf{0.4833} & 0.465 & 0.4505 \\
    LIR & 0.2833 & 0.3267 & 0.28 & 0.3367 & \textbf{0.395} & 0.3633 & 0.3 & 0.3264 \\
    LIR-2 & 0.34 & 0.36 & 0.34 & 0.38 & \textbf{0.44} & \textbf{0.44} & 0.38 & 0.3829 \\
    Single Influencer & \textbf{0.375} & 0.3583 & 0.3583 & 0.35 & 0.3583 & 0.3667 & \textbf{0.375} & 0.3631 \\
    Random & 0 & 0 & \textbf{0.01} & \textbf{0.01} & \textbf{0.01} & \textbf{0.01} & \textbf{0.01} & 0.0071 \\
    VoteRank & 0.3231 & 0.3477 & 0.3215 & \textbf{0.3569} & 0.3446 & 0.3446 & 0.3538 & 0.3418 \\
    \hline
    Overall & 0.2529 & 0.2646 & 0.2439 & 0.2634 & 0.2805 & \textbf{0.2855} & 0.2667 & 0.2654 \\
    \hline
    \end{tabular}
    \caption{Average accuracy over all top-$k$ algorithms for algorithms and link prediction metrics using $70\%$ training graphs from a $500$-node Erdös-Rényi graph.}
    \label{tab:rg_sms_acc_70}
\end{table*}

\begin{table*}[t]
    \centering
    \begin{tabular}{ | p{2.5cm} | p{1.7cm} |p{1.5cm} |p{1.2cm} |p{1.2cm} |p{1.2cm} |p{1.2cm} |p{1.7cm} |p{1.2cm} |}
    \hline
    Algorithm & Common Neighbors & Jaccard & Local Path & Quasi-Local RA & Quasi-Local RA-2 & RA-2 & Resource Allocation & Overall \\
    \hline
    Graph Coloring & 0.0233 & 0.0233 & \textbf{0.0600} & \textbf{0.0600} & \textbf{0.0600} & 0.0233 & 0.0233 & 0.0390 \\
    Joint Nomination & 0.0067 & 0.0117 & 0.0100 & 0.0133 & \textbf{0.0150} & 0.0133 & 0.0100 & 0.0114 \\
    k Highest & \textbf{0.6867} & 0.6783 & 0.5867 & 0.5833 & 0.5850 & \textbf{0.6867} & 0.6850 & 0.6417 \\
    LIR & 0.5217 & 0.5367 & 0.5067 & 0.5433 & \textbf{0.6033} & \textbf{0.6033} & 0.5483 & 0.5519 \\
    LIR-2 & 0.6400 & 0.6400 & 0.6400 & \textbf{0.6600} & \textbf{0.6600} & \textbf{0.6600} & \textbf{0.6600} & 0.6514 \\
    Single Influencer & \textbf{0.6167} & 0.6000 & 0.5583 & 0.5417 & 0.5417 & 0.6000 & 0.5917 & 0.5786 \\
    Random & 0.0000 & 0.0000 & \textbf{0.0100} & \textbf{0.0100} & \textbf{0.0100} & \textbf{0.0100} & 0.0000 & 0.0057 \\
    VoteRank & 0.6 & 0.62 & 0.5954 & 0.6354 & 0.6308 & 0.62 & 0.6277 & 0.6185 \\
    \hline
    Overall & 0.4338 & 0.4361 & 0.4149 & 0.4264 & 0.4345 & \textbf{0.4497} & 0.4411 & 0.4338 \\
    \hline
    \end{tabular}
    \caption{Average accuracy over all top-$k$ algorithms for algorithms and link prediction metrics using $90\%$ training graphs from a $500$-node Erdös-Rényi graph.}
    \label{tab:rg_sms_acc_90}
\end{table*}

\noindent \textbf{\textit{Mean Squared Error (MSE) Comparison.}} 
We averaged the MSE for all algorithms and their parameters over each prediction metric in both scenarios tested (see \autoref{tab:rg_sms_mse_70} and \autoref{tab:rg_sms_mse_90}, where the smallest value in each column appears in bold). In each case, the MSE was quite low, averaging around $0.11$ for complex contagion and $0.00014$ for simple contagion. It is interesting to note that in simple contagion, less variance was observed; this is likely due to the deterministic structure of our simplified influence model. However, even in the more realistic complex contagion scenario we also saw a relatively small mean squared error in the fraction of nodes infected, implying that overall the influence plots for the predicted graphs are quite close to that of the future graph.\\



\noindent \textbf{\textit{Accuracy Comparison.}} 
We averaged the accuracy for all algorithms over each centrality metric and prediction metric in both scenarios tested (see \autoref{tab:rg_sms_acc_70} and \autoref{tab:rg_sms_acc_90}, where the largest value(s) in each row appear in bold). In these tables, we can see the  overall predicting ability: the highest accuracies in the $70\%$ scenario are around $0.5$ (implying around half of the chosen vertices are the same on average) and are $0.7$ in the $90\%$ scenario. Therefore one has a high accuracy in the overall selection of influencers, meaning that our predicted graphs do not stray far from the true future graph. In particular, notice that there are a total of $500$ vertices in the graph; when chosen randomly there is an expected overlap of $\frac{5}{500} \cdot 5 = 0.05$ influencers from linearity of expectation, i.e. an accuracy of $0.0001$.

Interestingly, there are also noticeable differences between the overall accuracies of different algorithms. Overall, the selection of the $k$ highest-scoring centrality metrics performed best in each scenario, with VoteRank, LIR, LIR$-2$, and single influencer selection (taking the top scorer with respect to each centrality metric) also performing relatively well. Graph coloring and joint nomination do not perform well with respect to accuracy, however, which is to be expected. For the former, note that with the addition of our various predicted edges, the division of the graph into independent sets was likely changed drastically, which then throws off the accuracy of the nodes chosen in the predicted graphs, and for the latter notice that the algorithm itself relies on a semi-random selection, meaning similar selection of vertices is already quite unlikely. However, since we have weighted the probability distributions when selecting for joint nomination, we do see higher accuracy than expected. \\

\noindent \textbf{\textit{Algorithm Comparisons.}} 
We compared both the total overall performance differences by algorithm and their relative performances when also considering a prediction metric. 

\begin{figure}[h]
    \centering
    \begin{subfigure}{0.5\textwidth}
        \centering
        \includegraphics[width=\textwidth]{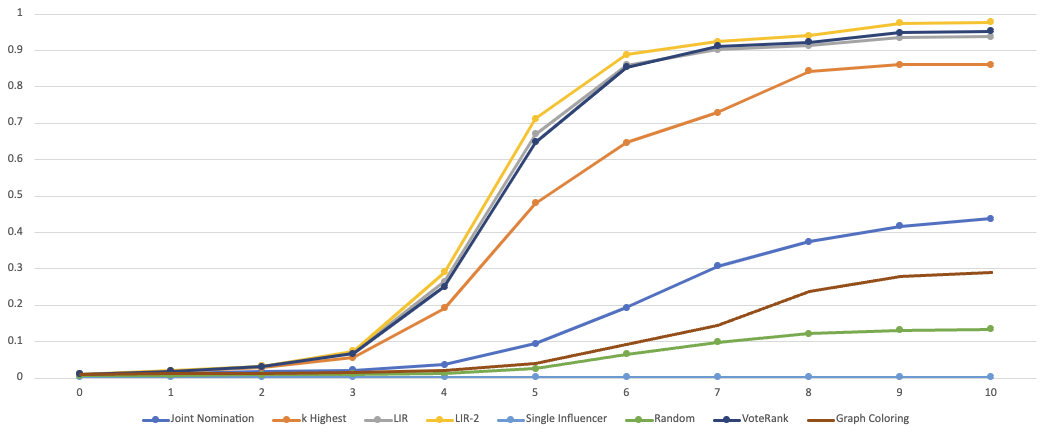}
        \caption{On complex contagion}
    \end{subfigure}
    \begin{subfigure}{0.5\textwidth}
        \centering
        \includegraphics[width=\textwidth]{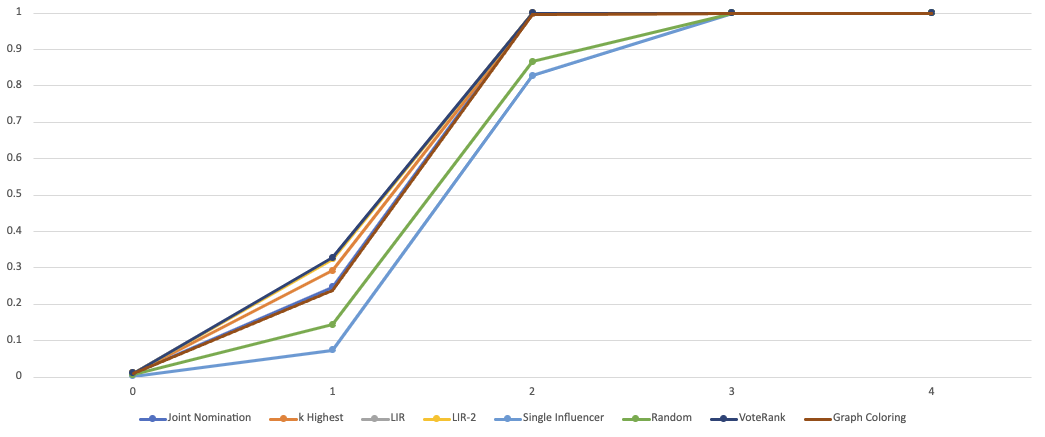}
        \caption{On simple contagion}
    \end{subfigure}
    \caption{Average performance of each algorithm over all graphs as fraction infected over time using $70\%$ training graphs from a $500$-node Erdös-Rényi graph.}
    \label{rg_70_alg_performance}
\end{figure}

\begin{figure}[h]
    \centering
    \begin{subfigure}[b]{0.5\textwidth}
        \centering
        \includegraphics[width=\textwidth]{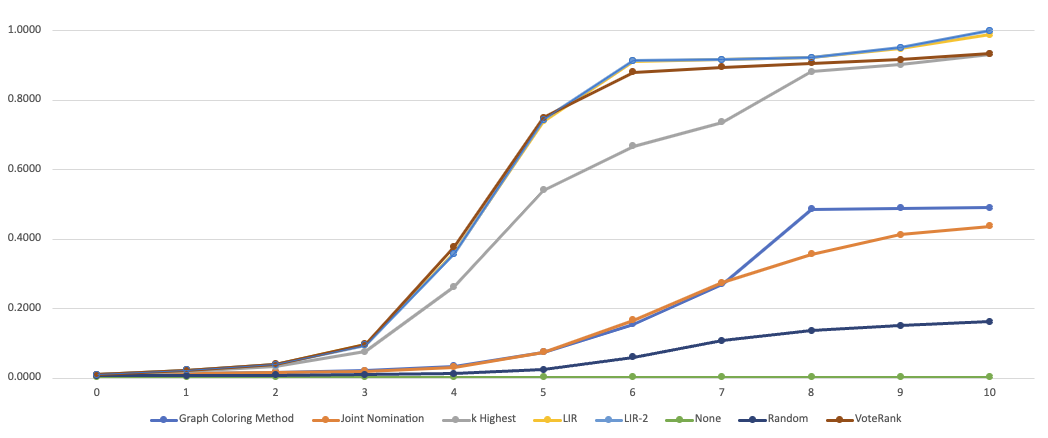}
        \caption{On complex contagion}
    \end{subfigure}
    \begin{subfigure}[b]{0.5\textwidth}
        \centering
        \includegraphics[width=\textwidth]{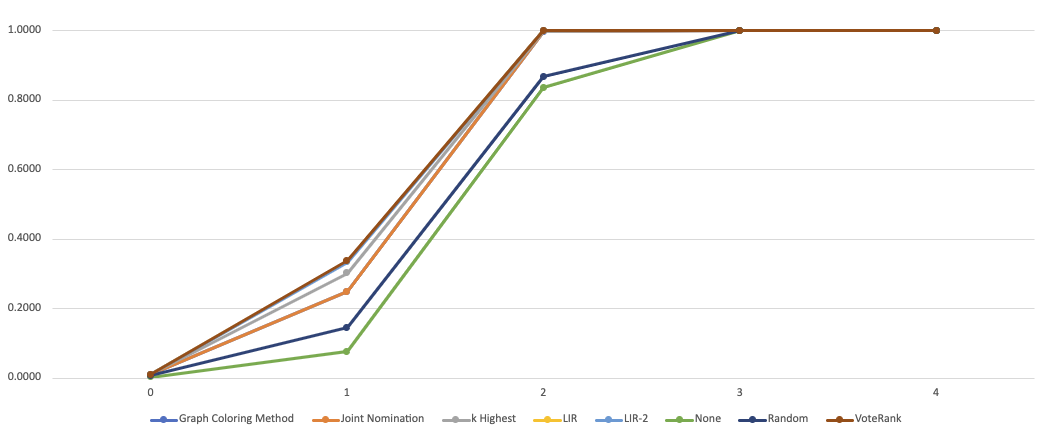}
        \caption{On simple contagion}
    \end{subfigure}
    \caption{Average performance of each algorithm over all graphs as fraction infected over time using $90\%$ training graphs from a $500$-node Erdös-Rényi graph.}
    \label{rg_90_alg_performance}
\end{figure}

We can see from \autoref{rg_70_alg_performance} and \autoref{rg_90_alg_performance} that LIR, LIR$-2$, and VoteRank are the best performers overall. Surprisingly, the $k$ highest algorithm is not too shabby either, surpassing joint nomination and graph coloring in each scenario. Possible explanations for this is first, that the purpose of joint nomination lies with disconnecting the graph rather than creating high influence spread, and second, that since some predicted graphs work better with graph coloring than others due to the aforementioned upsetting of the colorings, its average may have been lowered. As expected, random selection and single influencer selection (labeled as `None' in the figures) perform worst overall.

\begin{figure*}[h]
    \centering
    \begin{subfigure}[b]{0.45\textwidth}
        \centering
        \includegraphics[width=\textwidth]{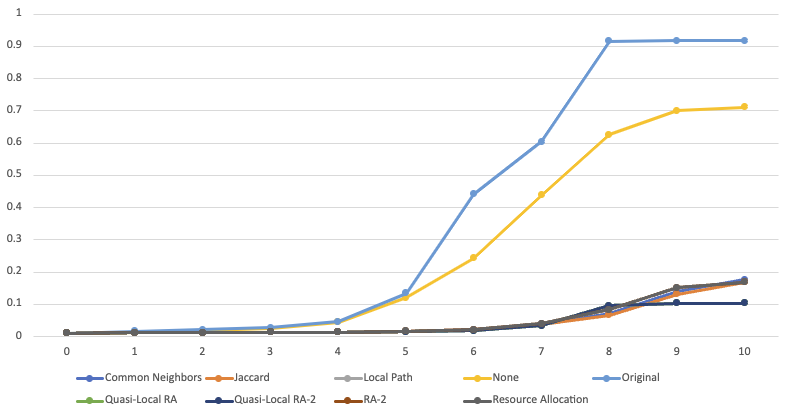}
        \caption{Graph Coloring}
    \end{subfigure}
    \hfill
    \begin{subfigure}[b]{0.45\textwidth}
        \centering
        \includegraphics[width=\textwidth]{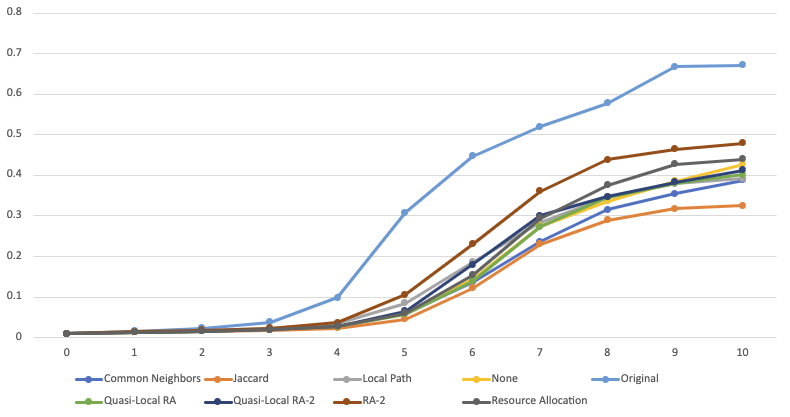}
        \caption{Joint Nomination}
    \end{subfigure}
    \hfill
    \begin{subfigure}[b]{0.45\textwidth}
        \centering
        \includegraphics[width=\textwidth]{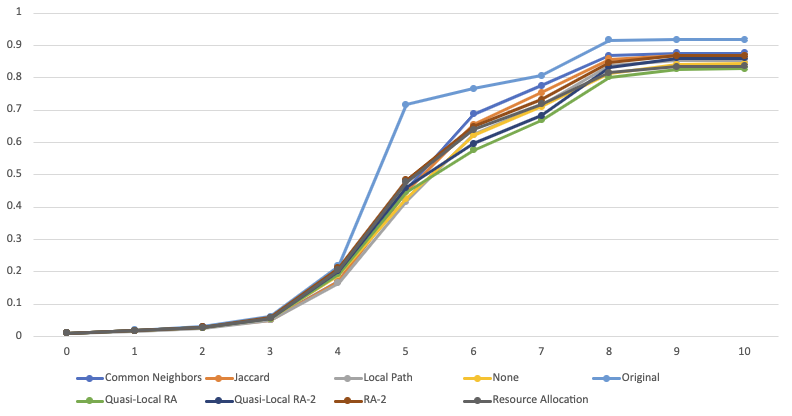}
        \caption{$k$ Highest}
    \end{subfigure}
    \hfill
    \begin{subfigure}[b]{0.45\textwidth}
        \centering
        \includegraphics[width=\textwidth]{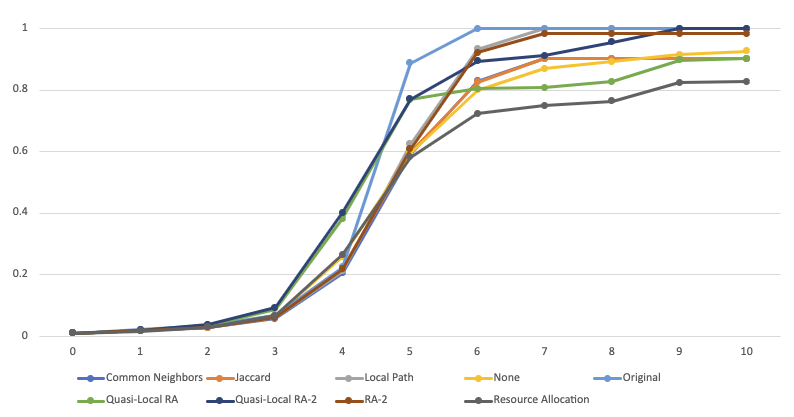}
        \caption{LIR}
    \end{subfigure}
    \hfill
    \begin{subfigure}[b]{0.45\textwidth}
        \centering
        \includegraphics[width=\textwidth]{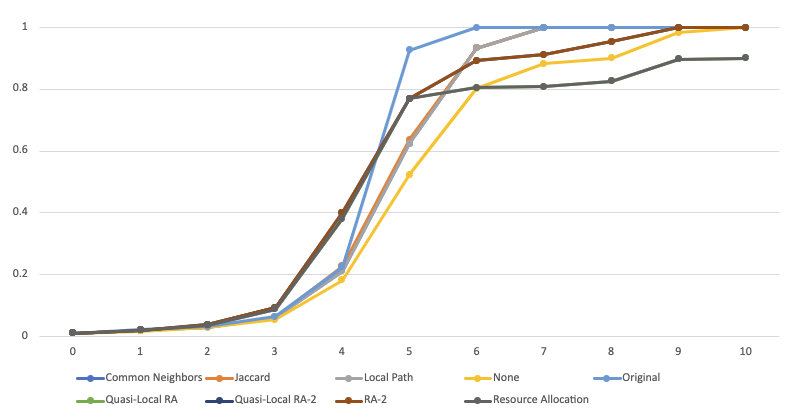}
        \caption{LIR$-2$}
    \end{subfigure}
    \hfill
    \begin{subfigure}[b]{0.45\textwidth}
        \centering
        \includegraphics[width=\textwidth]{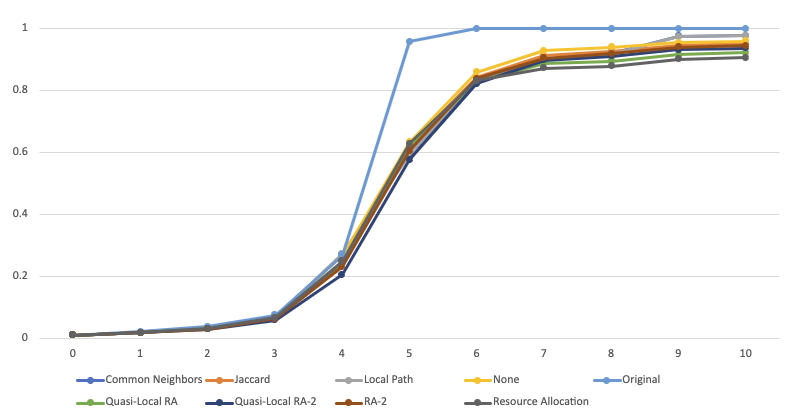}
        \caption{VoteRank}
    \end{subfigure}
    \caption{Average performance of each algorithm and prediction metric as fraction infected over time using $70\%$ training graphs from a $500$-node Erdös-Rényi graph.}
    \label{fig:rg_70_alg_sms}
\end{figure*}

\begin{figure*}[h]
    \centering
    \begin{subfigure}[b]{0.45\textwidth}
        \centering
        \includegraphics[width=\textwidth]{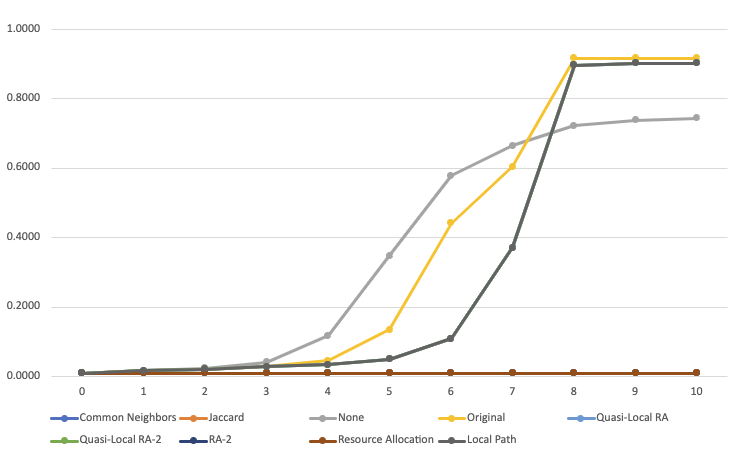}
        \caption{Graph Coloring}
    \end{subfigure}
    \hfill
    \begin{subfigure}[b]{0.45\textwidth}
        \centering
        \includegraphics[width=\textwidth]{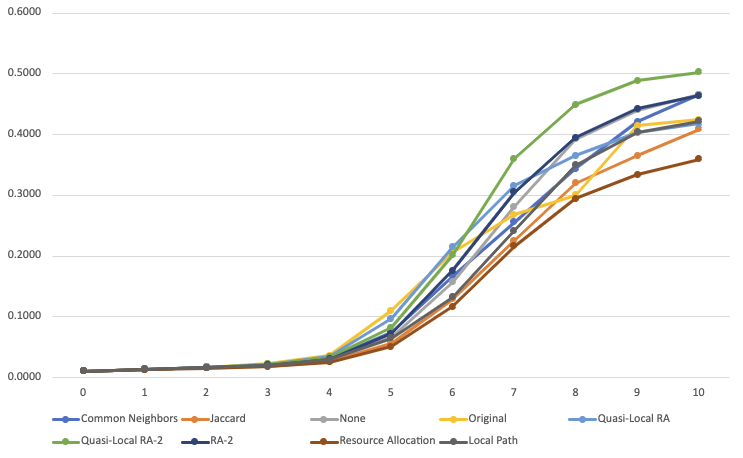}
        \caption{Joint Nomination}
    \end{subfigure}
    \hfill
    \begin{subfigure}[b]{0.45\textwidth}
        \centering
        \includegraphics[width=\textwidth]{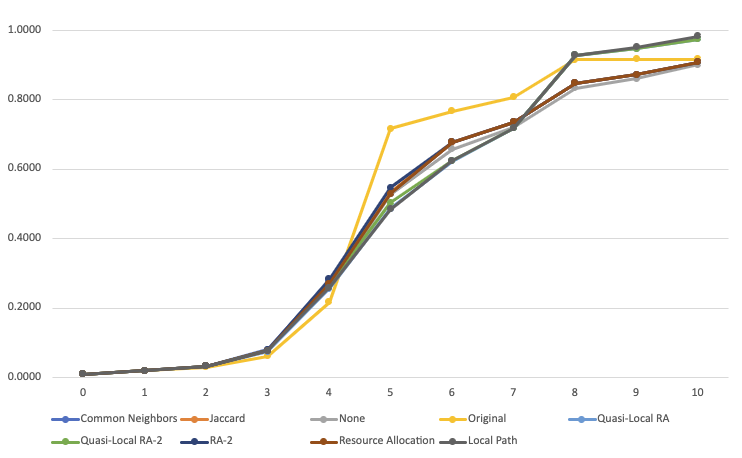}
        \caption{$k$ Highest}
        \label{subfig:rg_90_k_sms}
    \end{subfigure}
    \hfill
    \begin{subfigure}[b]{0.45\textwidth}
        \centering
        \includegraphics[width=\textwidth]{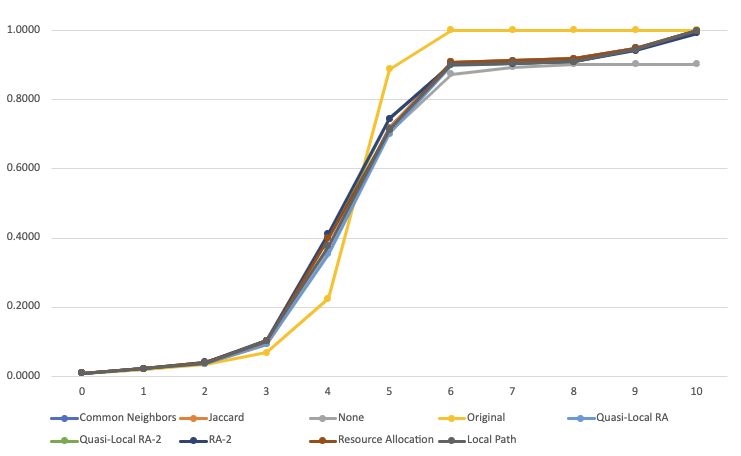}
        \caption{LIR}
    \end{subfigure}
    \hfill
    \begin{subfigure}[b]{0.45\textwidth}
        \centering
        \includegraphics[width=\textwidth]{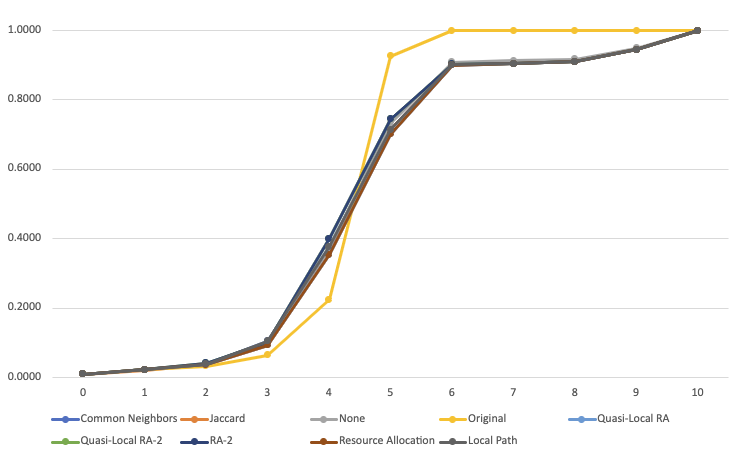}
        \caption{LIR$-2$}
    \end{subfigure}
    \hfill
    \begin{subfigure}[b]{0.45\textwidth}
        \centering
        \includegraphics[width=\textwidth]{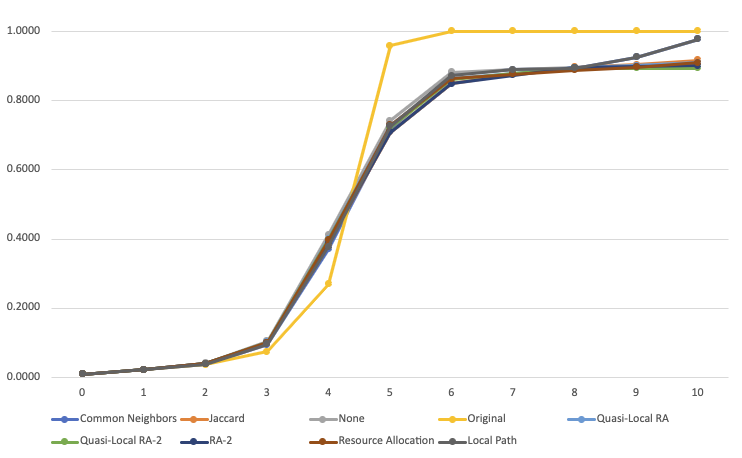}
        \caption{VoteRank}
    \end{subfigure}
    \caption{Average performance of each algorithm and prediction metric as fraction infected over time using $90\%$ training graphs from a $500$-node Erdös-Rényi graph.}
    \label{fig:rg_90_alg_sms}
\end{figure*}

When considering different prediction metrics, we will look only at data from the complex contagion models (see \autoref{fig:rg_70_alg_sms} and \autoref{fig:rg_90_alg_sms}), as the simple contagion data is very close together for each pair of algorithms and prediction metrics, making it less interesting. 
In most of the graphs, we can see that compared to `None,' the training graph itself, there are a substantial number of predicted graphs that perform consistently better (RA$-2$ and QR$-2$, for instance). Moreover, the predicted graphs all display influence approaching or, in \autoref{subfig:rg_90_k_sms}, even surpassing the averages for the original graph, suggesting an improvement in prediction when our Social Sphere Model is used.

\begin{figure*}
    \centering
    \includegraphics[width=1.0\textwidth]{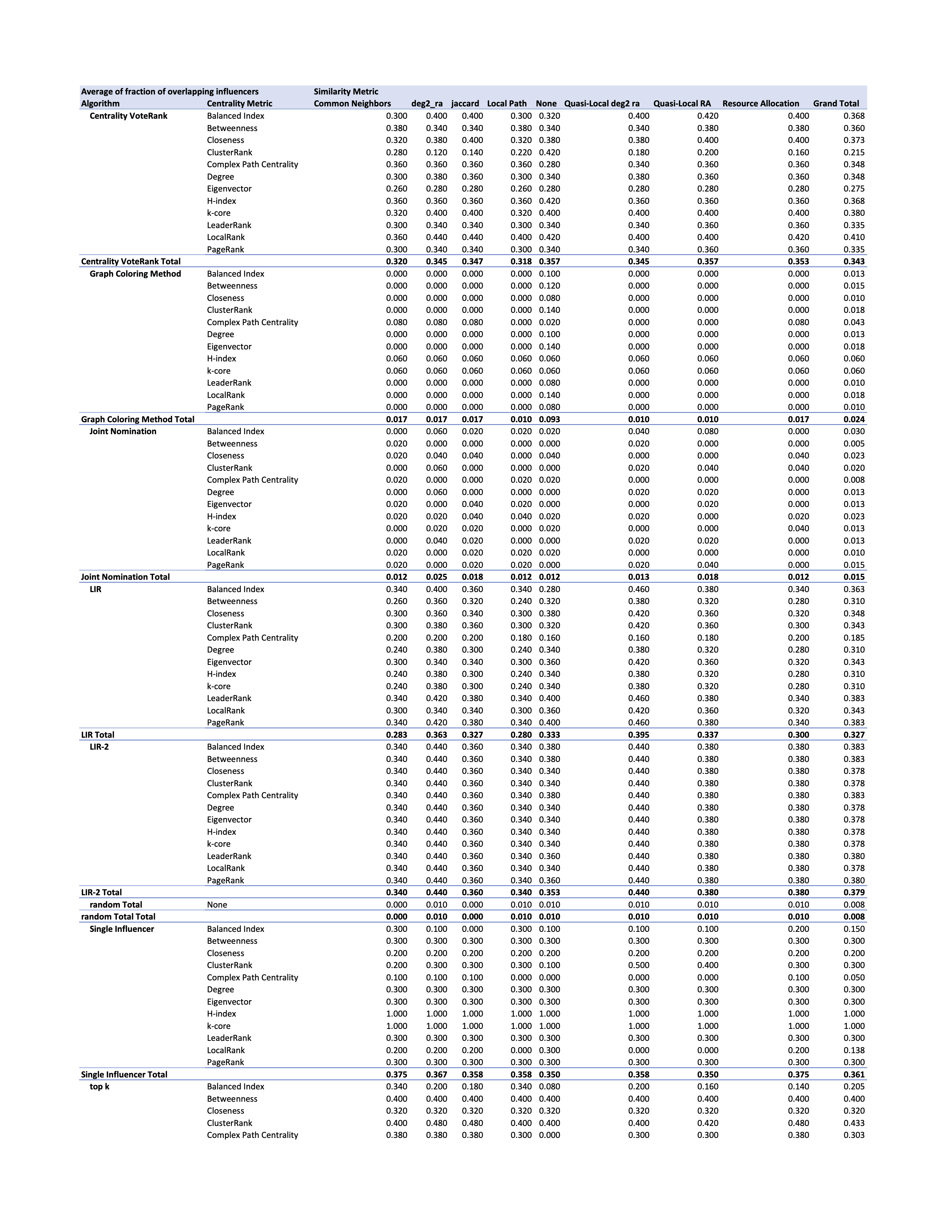}
    \caption{Average of accuracies for different algorithms on $70\%$ training graphs for the Erdös-Rényi random graph.}
    \label{fig:rg_70_sims}
\end{figure*}
\begin{figure*}
    \centering
    \includegraphics[width=1.0\textwidth]{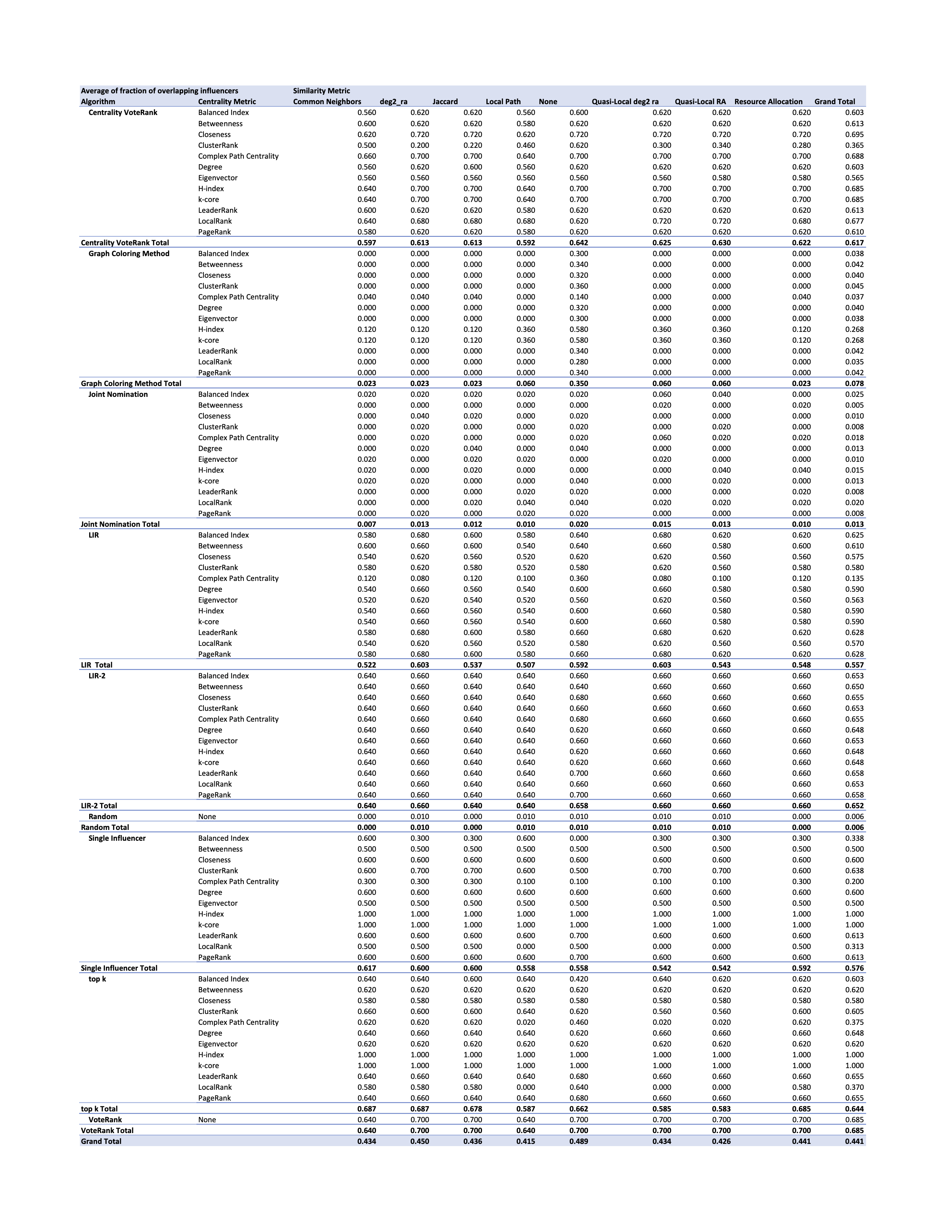}
    \label{fig:rg_90_sims}
    \caption{Average accuracies for different algorithms on $90\%$ training graphs for the Erdös-Rényi random graph.}
\end{figure*}

\section{Conclusion} \label{sec:conclusion}
In this research, we introduced the \textit{\textbf{Social Sphere Model}}, a novel algorithm which utilizes heuristic link prediction and influencer identification to forecast future influencers in weighted networks. Our comprehensive analyses on a complex random graph have demonstrated the algorithm's efficacy, particularly evident in the low mean squared error and relatively high accuracy, suggesting the Social Sphere Model's promising prediction capabilities for detecting future influencers, which is similar to Yanchenko et al. \cite{exante}'s findings that the influence of their model's selected nodes were comparable to that of the Oracle method.

By modifying and extending several existing influencer identification algorithms to incorporate centrality metrics, we found that the modified forms of VoteRank were comparable to the original form of VoteRank when using metrics such as betweenness. Our proposed similarity metric, RA-$2$, particularly excelled in both mean squared error and accuracy for the $90\%$ graphs. Moreover, our analysis of different link prediction metrics differing mostly in the degree of their denominators poses interesting questions around the optimal metric when applying the Social Sphere Model for different graphs and parameters.

The practical applications of our algorithm span a wide array of fields from marketing to epidemiology, offering valuable insights for strategic planning and information control. These include:
\begin{itemize} 
\item \textit{Viral marketing}: in the scenario mentioned in Section~\ref{sec:intro}, taking our algorithm over existing social media data and estimating $t$ from the expected time for production can assist in identifying potential future influencers in the network, giving more time (a valuable commodity) for negotiation, scouting, and strategizing. The relatively short time complexities of heuristic models are also sometimes valuable in these scenarios. As an analogy, if one wanted to understand how admissions would look like in a {\it university program for influencers}, our algorithm would allow one to evaluate the potential of an influencer or group of influencers in the wider network.
\item \textit{Better Network Approximation}: The already-documented \cite{crimeprediction} utility of link prediction in approximating incomplete network data further underscores the significance of our approach, especially considering the typically resource-intensive nature of complete data collection. 
\item \textit{Historical Consideration}: Our results show potential in identifying hidden or overlooked influencers, which presents a dynamic aspect that can enhance performance and integrate more information in the search for influencers.
\end{itemize}

In particular, our model differentiates from other ones by making use of relatively low time complexity metrics and algorithms as opposed to neural networks, the training of which may be infeasible for smaller businesses that might seek to calculate future influencers. The simplicity and efficiency of it can thus allow for quick identification of influencers and thus better networking. Finally, it would be of much interest to study the implications of the Social Sphere Model more empirically on various synthetic networks and real-world datasets such as Yelp, Wikipedia, Coauthorship Network, and Twitter while investigating properties like varying clustering coefficient or diameter values.

Finally, whilst this paper focuses on the theoretical formulation of the Social Sphere Model, including its
integration of similarity-based link prediction, and evaluation of the model on random graphs, a
companion work will examine the empirical performance and practical applications of the model
on empirical datasets. Our second study, titled "Empirical Evaluation of the Social Sphere Model
on Influencer Identification on Social Networks," we evaluate the model's behavior in a practice
setting on author collaboration networks, and its predictive ability in identifying latent influencers,
and explore the application for information diffusion and digital marketing strategies. The two
works put together aim to provide a comprehensive investigation of the Social Sphere Model’s
capabilities from a theoretical lens and in real-world datasets.

\section*{Acknowledgments}

We extend our gratitude to the MIT PRIMES-USA program for their invaluable support and guidance throughout this research. The research of LPS is partially supported by  NSF FRG Award DMS- 2152107, a Simons Fellowship and NSF CAREER Award DMS 1749013. Part of the work appearing here was done whilst LPS was a Visiting Fellow at All Souls College, Oxford.

\printcredits

\pagebreak
\bibliographystyle{cas-model2-names}

\bibliography{cas-refs}
\end{document}